\documentclass{article}  

\usepackage{booktabs}
\usepackage{amsmath}
\usepackage{authblk}
\usepackage{graphicx}
\usepackage{abstract}
\usepackage{amssymb}
\usepackage{subcaption}
\usepackage{fancyhdr}
\usepackage{hyperref}
\usepackage{color}
\usepackage[round]{natbib}

\title{A Statistical Mechanical Approach for the Parametrization of the Coupling in a Fast-Slow System}

\author[1,2]{Gabriele Vissio}
\author[2,3,4]{Valerio Lucarini}

\affil[1]{International Max Planck Research School on Earth System Modelling, Hamburg, Germany}
\affil[2]{CEN, Meteorological Institute, University of Hamburg, Hamburg, Germany}
\affil[3]{Department of Mathematics and Statistics, University of Reading, Reading, UK}
\affil[4]{Walker Institute for Climate System Research, University of Reading, Reading, UK}
\begin{document}
\maketitle
\begin{onecolabstract}
Constructing accurate, flexible, and efficient parametrizations is one of the great challenges in the numerical modelling of geophysical fluids. We consider here the simple yet paradigmatic case of a Lorenz 84 model forced by a Lorenz 63 model and derive a parametrization using a recently developed statistical mechanical methodology based on the Ruelle response theory. We derive an expression for the deterministic and the stochastic component of the parametrization and we show that the approach allows for dealing seamlessly with the case of the Lorenz 63 being a fast as well as a slow forcing compared to the characteristic time scales of the Lorenz 84 model. We test our results using both standard metrics based on the moments of the variables of interest as well as Wasserstein distance between the projected measure of the original system on the Lorenz 84 model variables and the measure of the parametrized one. By testing our methods on reduced phase spaces obtained by projection, we find support to the idea that comparisons based on the Wasserstein distance might be of relevance in many applications despite the curse of dimensionality.
\end{onecolabstract}

\section{Introduction}
The climate is a forced and dissipative system featuring variability on a large range of spatial and temporal scales, as a result of many complex and coupled dynamical processes inside it \citep{Peixoto1992,Lucarini2014,Ghil2015}. Numerical models are able to resolve explicitly only a relatively small range of such scales. In particular, it is crucial to derive efficient and accurate ways to surrogate the effect of dynamical processes occurring on small spatial and temporal (so-called subgrid) scales, which are not explicitly resolved (e.g. because of enormous computational cost or theoretical uncertainties), on the scales of interest, explicitly resolved by the model. The operation of constructing so-called parametrizations is key to the development of geophysical fluid dynamical models and stimulates the investigation of the fundamental laws defining the multiscale properties of the coupled atmosphere-ocean dynamics \citep{Uboldi2015,Vannitsem2016}. Traditionally, the development of parametrizations boiled down to deriving deterministic empirical laws able to describe the effect of the small scale dynamical processes. More recently, it has become apparent the need to include stochastic terms able to provide a theoretically more coherent representation of such effects and, at practical level, an improved skill \citep{Palmer2008,Franzke2015,Berner2017}.

A first way to derive or at least justify the need for stochastic parametrizations comes from homogenization theory \citep{Pavliotis2008}, which leads to constructing an approximate representation of the impact of the fast scales on the slow variables as the sum of two terms, a mean field term and a white noise term. Such an approach suffers from the fact that one has to take the rather nonphysical hypothesis that an infinite time scale separation exists between the fast and the slow scale. As the climate is a multiscale system, such a methodology is a bit problematic to adopt.

\cite{Mori1974} and \cite{Zwanzig1960,Zwanzig1961} analyzed, in the context of statistical mechanics, the related problem of studying how one can project out the effect of a group of variables, with the goal of constructing effective evolution equations for a subset of variables of interest. They discovered that, in general, one can surrogate such an operation of coarse graining by adding to the equations of motion of the variables of interest three extra terms, a deterministic term, a stochastic forcing and a memory term. The memory term defines a non-markovian contribution where the past states of the variables of interest enter the evolution equation. In the limit of infinite time scale separation such last term tends to zero, whilst the random forcing approaches the form of a (in general, multiplicative) white noise.

The triad of terms - deterministic, stochastic and non-markovian - was also found by \cite{Wouters2012,Wouters2013,Wouters2016}, who proposed a method for constructing parametrizations based on the Ruelle response theory \citep{Ruelle1998,Ruelle2009}. They interpreted the coupling between the variables of interest and those one wants to parametrize as a weak perturbation of the otherwise unperturbed uncoupled slow and fast systems. A useful feature of this methodology is that it can be applied on a wide variety of coupled system not necessarily scale separated, extending its applications beyond (yet including) multiscale models. The parametrizations obtained along these lines match the result of the perturbative expansion of the projection operator introduced by Mori and Zwanzig.

This method has already been successfully tested in the works of \cite{Wouters2016a,Demaeyer2017,Vissio2018}.

Conceptually similar results have been found through bottom up, data driven approaches, by \cite{Kravtsov2005,Chekroun2015a,Chekroun2015b,Kondrashov2015}. Specifically, \cite{Kravtsov2005} constructed effective models from climatic time series through an extension to the non linear case of the multilevel linear regressive method, while \cite{Kondrashov2015} showed how non-markovian data-driven parametrizations emerge naturally when we consider partial observations from a large-dimensional system.

Even when a parametrization is efficient enough to represent subgrid phenomena with the desired precision, problems arise when it comes to dealing with scale adaptivity. Re-tuning the parametrization to a new set of conditions usually means running again long simulations, adding further computational costs. For this reason the development of a scale adaptive parametrization is considered to be a central task in geosciences \citep{Arakawa2011,Park2014,Sakradzija2016}. In a previous paper, the authors demonstrated the scale adaptivity of the WL approach by testing it in a mildly modified version of the Lorenz 96 model \citep{Lorenz1996}. A further degree of flexibility of this approach has been explored in another recent publication \citep{Lucarini2017}.

In this paper, we wish to apply the WL parametrization to a simple dynamical system constructed using as system of interest Lorenz 84 \citep{Lorenz1984} and as a forcing system Lorenz 63 \citep{Lorenz1963}, where the latter influences the evolution of the former thorough a linear coupling and will be appropriately parametrized. This model, already used by \citep{Bodai2011}, will be drastically changed within the paper modifying the value of the time scale separation to switch the roles of slow and fast scale systems between the two models. We wish to extend what studied in \cite{Vissio2018} by focusing on doing a systematic comparison of the properties of the projected measure of the original coupled system on the subspace spanned by the variables of the Lorenz 84 model with the actual measure of the parametrized model. We will look into the properties of the first two moments of the variables and will also study specifically the Wasserstein distance \citep{Villani2009} between the two measures, which allows for a comprehensive evaluation of how different the attractors are. The Wasserstein distance has been proposed by \cite{Ghil2015} as a tool for studying the climate variability and response to forcings, and applied by \cite{Robin2017} in a simplified setting.

In Section \ref{models} we describe thoroughly the single models and the full coupled model, while in Section \ref{WLP} we summarily review Wouters-Lucarini's parametrization and its application to the Lorenz 84-Lorenz 63 coupled model. Section \ref{WasDis} is dedicated to expose the Wasserstein distance, a useful tool to ascertain quantitavely that the attractor drew by the system subject to the parametrization is similar to the one outlined by the full system. Section \ref{ParFast} shows the results of the tests performed. In the last Section we draw conclusions on the study undertaken.

\section{Models} \label{models}

\subsection{Lorenz 84}
The Lorenz 84 model \citep{Lorenz1984} provides an extremely simplified representation of the large scale atmospheric circulation:
\begin{align} 
\frac{dX}{dt}&=-Y^2-Z^2-aX+aF_0 , \label{L84X} \\
\frac{dY}{dt}&=XY-bXZ-Y+G , \label{L84Y} \\
\frac{dZ}{dt}&=XZ+bXY-Z . \label{L84Z}
\end{align}
where the variable $X$ describes the intensity of the westerlies, while the variables $Y$ and $Z$ correspond to the two phases of the planetary waves responsible for the meridional heat transport.
\newline Thus, Eq.\eqref{L84X} describes the evolution of the westerlies, subject to the external forcing $F_0$, dampened both from the linear term $-aX$ and from nonlinear interaction with the eddies $-Y^2$ and $-Z^2$. This interaction amplifies the eddies through the terms $XY$ and $XZ$ in Eqs.\eqref{L84Y}-\eqref{L84Z}, eddies displaced by the action of the westerlies through the terms $-bXZ$ and $bXY$. The constant $b$ regulates the relative rapidity between diplacements and amplifications. In Eqs. \eqref{L84Y}-\eqref{L84Z} we can, as in Eq.\eqref{L84X}, see a linear dissipation, whilst the symmetry between the two equations is broken by the external forcing $G$.

\subsection{Lorenz 63}
Lorenz 63 is probably the most iconic chaotic dynamical system \citep{Saltzman1962,Lorenz1963,Ott1993} and was developed from Navier-Stokes and thermal diffusion equations (see e.g. \cite{Hilborn2000} for a complete, yet simple, derivation of the model) to describe through a simple dynamical system the evolution of three modes corresponding to large scale motions and temperature modulations in the Rayleigh-B\'enard convection framework. The three equations are:
\begin{align} 
\frac{dx}{dt}&=\sigma (y-x) , \label{L63x} \\
\frac{dy}{dt}&=\rho x-y-xz , \label{L63y} \\
\frac{dz}{dt}&=-\beta z+xy , \label{L63z}
\end{align}
where $x$, $y$ and $z$ are proportional, respectively, to the intensity of the convective motion (being Eq.\eqref{L63x} obtained by the Navier-Stokes equation expressed for a streamfunction), to the difference between temperatures of upward and downward fluid flows and to the difference of the temperature in the center of a convective cell with respect to a linear profile (since Eqs.\eqref{L63y}-\eqref{L63z} derive from thermal diffusion equation).
\newline $\sigma$, $\rho$ and $\beta$ are constants which depend on kinematic viscosity, thermal conductivity, depth of the fluid, gravity acceleration, thermal expansion coefficient; specifically, $\sigma$ is also known as \textit{Prandtl Number}.

\subsection{Coupled model} \label{Coupled model}
The full model used in this paper, proposed by \cite{Bodai2011}, is constructed by coupling the two low-order models introduced before as follows. The Lorenz 63 system acts as a forcing for the Lorenz 84 system, which represents the dynamics of interest. The dynamics of the two systems has a time scale separation given by the factor $\tau$ and can be written as follows:
\begin{align} 
\frac{dX}{dt}&=-Y^2-Z^2-aX+a(F_0+hx) , \label{L84XC} \\
\frac{dY}{dt}&=XY-bXZ-Y+G , \label{L84YC} \\
\frac{dZ}{dt}&=XZ+bXY-Z , \label{L84ZC} \\
\frac{dx}{dt}&=\tau \sigma (y-x) , \label{L63xC} \\
\frac{dy}{dt}&=\tau (\rho x-y-xz) , \label{L63yC} \\
\frac{dz}{dt}&=\tau (-\beta z+xy) . \label{L63zC}
\end{align}

It is important to underline that the coupling between the Lorenz 84 and the Lorenz 63 is uni-directional: the latter model affects the former and, acts as an external forcing, with no feedback acting the other way around.

In what follows, the parameters are chosen from the common setting: $a=0.25$, $b=4$, $\sigma=10$, $\rho=28$, $\beta=8/3$; the two forcings are set as $F_0=8$, namely the winter and more chaotic condition, and $G=1$. $h$ is a modulation coefficient which defines the coupling strength and it is chosen as $0.25$ to provide a stochastic forcing between two and four orders of magnitude smaller (on average) than the tendencies of the $X$ variable.
\newline All Lorenz models have a time unit equal to $5$ days, and the time scale separation $\tau$ regulates the time scale of the forcing provided by Lorenz 63. In case of $\tau>1$, this model can be seen, for example, as the action of small scale convective motions which directly influences the large scale westerly winds; in this framework $\tau=5$ corresponds to a daily fluctuation. On the other hand, $\tau<1$ implies a forcing on longer time scales than Lorenz 84, e.g. an orbital effect; in this case we can have, for example, monthly ($\tau=\frac{1}{6}$) or annual ($\tau=\frac{1}{73}$) influences.

Henceforth, we will refer to the standard Lorenz 84 as \textit{uncoupled model}, whilst the Lorenz 84 subject to the coupling with the Lorenz 63 will be called \textit{coupled model}.

\section{Wouters Lucarini's parametrization} \label{WLP}
\cite{Wouters2012,Wouters2013,Wouters2016} presented a top-down method suitable for constructing parametrizations for chaotic dynamical systems in the form:
\begin{align}
  \frac{dK}{dt}&=F_K(K)+\epsilon\Psi_K(K,J), \label{perturbedX}\\
  \frac{dJ}{dt}&=F_J(J)+\epsilon\Psi_J(K,J), \label{perturbedY}
\end{align}
where the $K$ is the variable we are interested in and the $J$ is the variable we want to parametrize. The coefficient $\epsilon$ regulates the magnitude of the alterations to the evolution of the variable induced by the couplings.
\newline The parametrization is obtained assuming the chaotic hypothesis and applying Ruelle response theory \citep{Ruelle1998,Ruelle2009}; the coupling in \eqref{perturbedX} is approximated, up to the second order, by three terms: the first order consists in a deterministic term, while the second order includes a stochastic forcing and a non-markovian term. The general form of the parametrization (e.g. \cite{Vissio2018}) is:
\begin{equation} 
  \frac{dK}{dt}=F_K(K)+\epsilon D(K)+\epsilon S\{K\}+\epsilon^2 M\{K\} , \label{wl_par}
\end{equation}
where $D$, $S$ and $M$ indicate, respectively, deterministic, stochastic and memory term. A useful outcome of the linear response theory involved consists in the equations needed to calculate the three terms: since the couplings are seen as perturbation of an otherwise unperturbed system, those terms must be calculated considering the statistical properties of the unperturbed equations
\begin{align}
  \frac{dK}{dt}&=F_K(K) , \label{unperturbedX}\\
  \frac{dJ}{dt}&=F_J(J) . \label{unperturbedY}
\end{align}
The numerical integration of Eqs.\eqref{unperturbedX}-\eqref{unperturbedY} may allow to use less computational resources with respect to Eqs.\eqref{perturbedX}-\eqref{perturbedY}, particularly in the case of multiscale systems.

\subsection{Constructing the parametrization}

The coupling strength $\epsilon$, shown in Eqs.\eqref{perturbedX}-\eqref{perturbedY} and in Eq.\eqref{wl_par}, assumes the value $\epsilon=ah$.

The deterministic term $D$ in Eq.\eqref{wl_par} is a measure of the average impact of the coupling on the $K$ dynamics and assumes the form:
\begin{equation}
  D(K)=\rho_{0,J}(\Psi_K(J))= \lim_{T\to\infty}\frac{1}{T}\int_0^T \Psi_K(J)d\tau= \lim_{T\to\infty}\frac{1}{T}\int_0^T x(\tau)d\tau , \label{eq:Dterm}
\end{equation}
where we have used the expression of the coupling given in Eq.\ref{L84XC} and we have computed the ensemble average as time average on the ergodic measure.

Since the coupling shown in Eq.\eqref{L84XC} depends only on one of the variables (in this case the $x$) of the system we want to parametrize, the stochastic term can be written as
\begin{equation} 
  S\{K\}=\sigma(t) ,  \label{stochastic}
\end{equation}
where
\begin{equation} \label{eq:fluctuationterm}
\begin{split}
  R(t)=\langle\sigma(0),\sigma(t)\rangle&=\rho_{0,J}((\Psi_K(J)-D)(\Psi_K(f^{t}(J))-D)) ,\\
 &=\rho_{0,x}((x(0)-D)(x(t)-D)) ,\\
  \langle\sigma(t)\rangle&=0 .
\end{split}
\end{equation}
As discussed in \cite{Wouters2012,Wouters2013,Vissio2018}, for more complex couplings the stochastic terms assumes the form of a multiplicative noise. We have used the software package ARFIT \citep{Neumaier2001,Schneider2001} to construct time series of noise with the desired properties defined by Eqs.\ref{eq:fluctuationterm}.

The last term in Eq.\ref{wl_par} is the non-markovian contribution to the parametrization and can be written as follows:
\begin{equation} \label{eq:m3}
  M\{K\}=\int_0^\infty h(t_2,K(t-t_2))dt_2 ,
\end{equation}
where
\begin{equation} \label{eq:new_h}
  h(t_2,K)=\Psi_J(K)\rho_{0,J}(\partial_J\Psi_K(f^{t_2}(J)))=0 \cdot \rho_{0,x}(\partial_x x(f^{t_2}(x))).
\end{equation}
As discussed in Section \ref{Coupled model}, the evolution of the variables of the Lorenz 63 model are independent of the state of the variables corresponding to the Lorenz 84 model. As a result, the first factor on the r.h.s. of Eq.\ref{eq:new_h} vanishes, so that the parametrization we derive is fully markovian.

After the implementation of Wouters-Lucarini's procedure, Eq.\eqref{L84XC} will be parametrized as
\begin{equation} 
\frac{dX}{dt}=-Y^2-Z^2-aX+a[F_0+h(D+S)] ; \label{L84XP2nd}
\end{equation}
Eq.\eqref{L84XP2nd}, together with Eqs.\eqref{L84YC}-\eqref{L84ZC}, will be henceforth indicated as \textit{second order parametrization}, whilst the same equations without the stochastic term (therefore comprehending the first order, deterministic term only), namely
\begin{equation} 
\frac{dX}{dt}=-Y^2-Z^2-aX+a[F_0+hD] , \label{L84XP1st}
\end{equation}
will be called \textit{first order parametrization}.

\section{Wasserstein Distance} \label{WasDis}
We wish to  assess how well a parametrization allows to reproduce the statistical properties of the full coupled system. At this regard, it seems relevant to quantify to what extent the projected invariant measure of the full coupled model on the variables of interest differs from the invariant measures of the surrogate models containing the parametrization. In order to evaluate how much such measures differ, we resort to considering their Wasserstein distance \citep{Villani2009}. Such a distance quantifies the minimum "effort" in morphing one measure into the other, and was originally introduced by \cite{Monge1781}, somewhat unsurprisingly, to study problems of military relevance, and later improved by \cite{Kantorovich1942}.
 
Starting from two distinct spatial distribution of points, described by the measure $\mu$ and $\nu$, we can define the optimal transport cost \citep{Villani2009} as the minimum cost to move the set $\mu$ into $\nu$:
\begin{equation} 
C(\mu,\nu)=\inf_{\pi \in \Pi(\mu,\nu)} \int c(x,y) d \pi(x,y) , \label{opttrcost}
\end{equation}
where $c(x,y)$ is the cost for transporting one unit of mass from $x$ to $y$. The function $C(\mu,\nu)$ in Eq.\eqref{opttrcost} is called Kantorovich-Rubinstein distance. In the rest of the paper, we will consider the Wasserstein distance of order 2:
\begin{equation} 
W_2(\mu,\nu)= \left\lbrace  \inf_{\pi \in \Pi(\mu,\nu)} \int [d(x,y)]^2 d \pi(x,y) \right\rbrace ^ {\frac{1}{2}} . \label{wd2continous}
\end{equation}

We can define the Wasserstein distance also in the case of two discrete distributions
\begin{align} 
\mu&=\sum\limits_{i=1}^n \mu_i \delta_{x_i} ,\\
\nu&=\sum\limits_{i=1}^n \nu_i \delta_{y_i} ,
\end{align}
where $x_i$ and $y_i$ represent the location of the different points, which mass is given, respectively, by $\mu_i$ and $\nu_i$. Recalling the definition of Euclidean distance
\begin{equation} 
d(\mu,\nu)=\left[ \sum\limits_{i=1}^n (x_i-y_i)^2 \right] ^ {\frac{1}{2}} ,
\end{equation}
and representing with $\gamma_{ij}$ the fraction of mass transported from $x_i$ to $x_j$, we can construct the order 2 Wasserstein distance for discrete distributions as follows:
\begin{equation} 
W_2(\mu,\nu)= \left\lbrace \inf_{\gamma_{ij}} \sum\limits_{i,j} \gamma_{ij} [d(x_i,y_j)]^2 \right\rbrace ^ {\frac{1}{2}} . \label{wd2discrete}
\end{equation}

This latter definition of Wasserstein distance has already been proven effective \citep{Robin2017} for providing a quantitative measurement of the difference between the snapshot attractors of the Lorenz 84 system in the instance of summer and winter forcings.

Hereby we propose to further assess the reliability of WL stochastic parametrization through the application of the Wasserstein distance. Nevertheless, since the numerical computations for optimal transport through linear programming theory are not cheap, a new approach is required. In order to accomplish it, we perform a standard Ulam discretization of the measure supported on the attractor \citep{Ulam1964,Tantet2018}. By coarse-graining on a set of cubes with constant sides across the phase space. We will discuss below the impact of changing the sides of such cubes.

The coordinates of the cubes will then be equal to the location $x_i$, while the correspondent densities of the points are used to define $\gamma_{ij}$; finally, we exclude from the subsequent calculation all the grid boxes containing no points at all.
\newline Our calculations are performed using a modified version of the software for Matlab written by  Gabriel Peyr\'e and made available at \url{http://www.numerical-tours.com/matlab/optimaltransp_1_linprog/}, conveniently modified to include the subdivision of the phase space in cubes and the assignment of correspondent density to those cubes.

\section{Parametrizing the Coupling with the Lorenz 63 Model} \label{ParFast}
In this section we show the results corresponding to the case $\tau=5$. Therefore, Lorenz 84 and Lorenz 63 are seen as, respectively, the slow and the fast dynamical systems.

\subsection{Qualitative Analysis} \label{QualAnal}
\begin{figure*}
\begin{subfigure}{.5\textwidth}
  \centering
  \includegraphics[width=\linewidth]{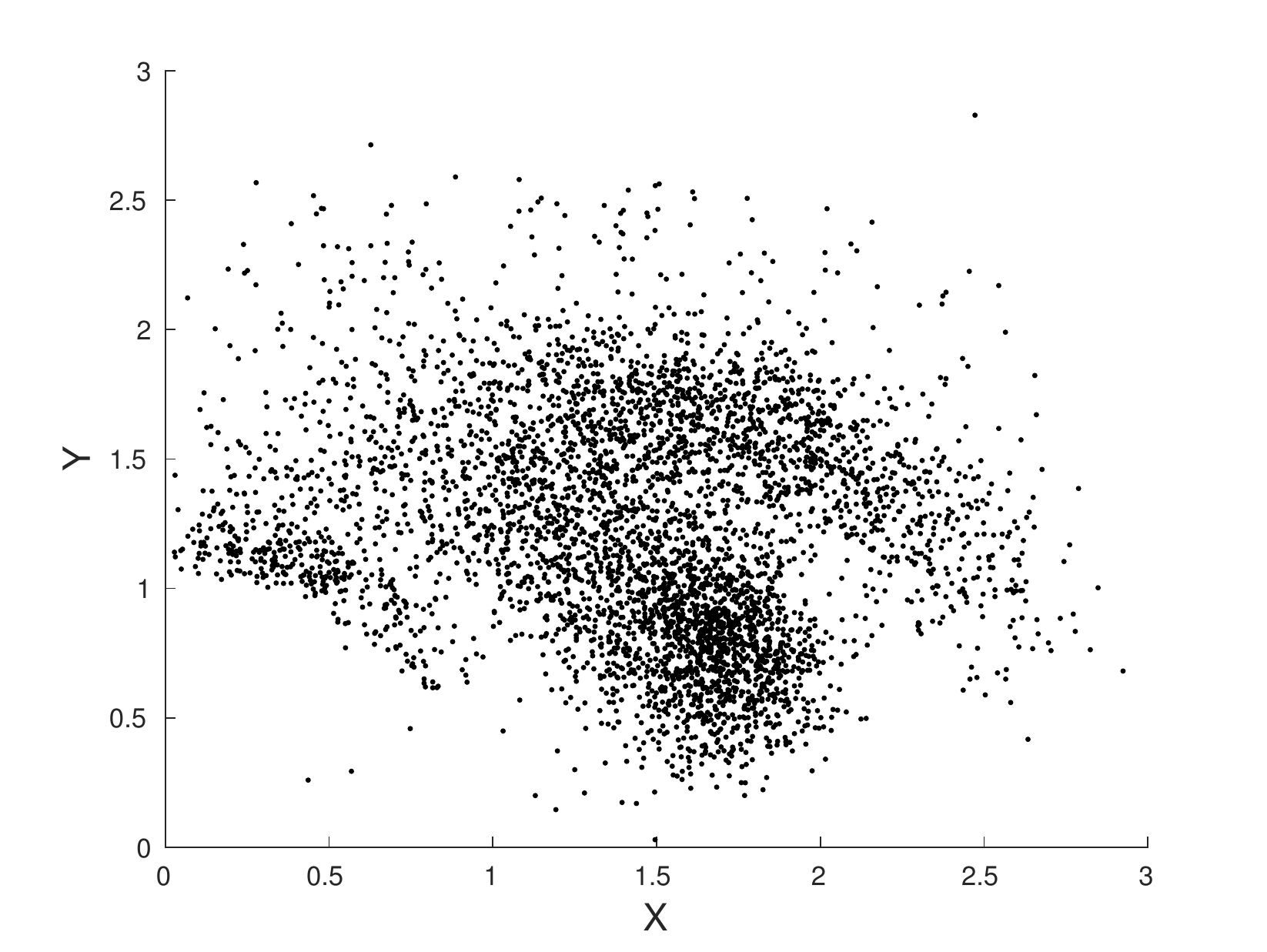}
  \caption{}
\end{subfigure}%
\begin{subfigure}{.5\textwidth}
  \includegraphics[width=\linewidth]{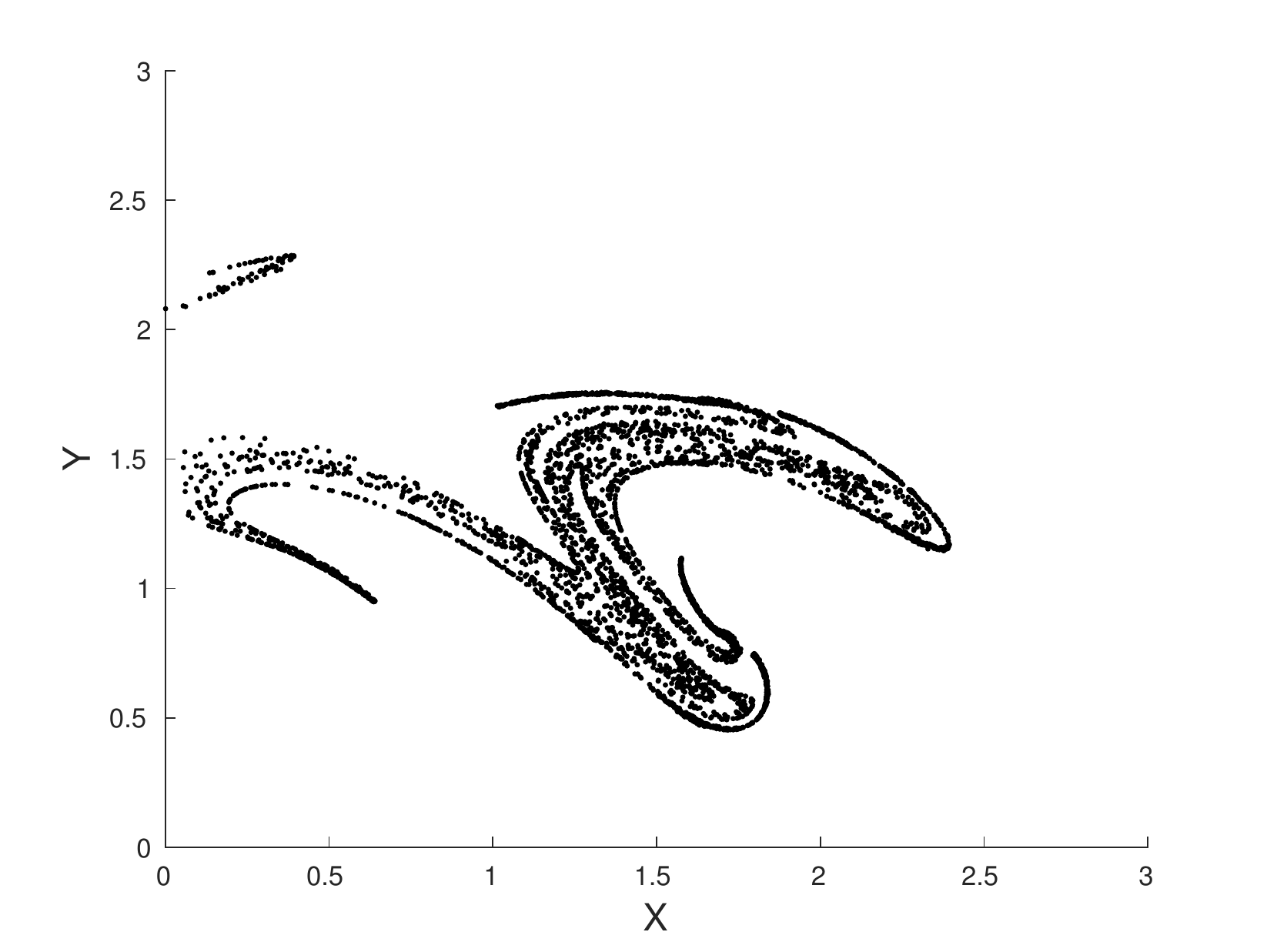}
  \caption{}
\end{subfigure}\\
\begin{subfigure}{.5\textwidth}
  \centering
  \includegraphics[width=\linewidth]{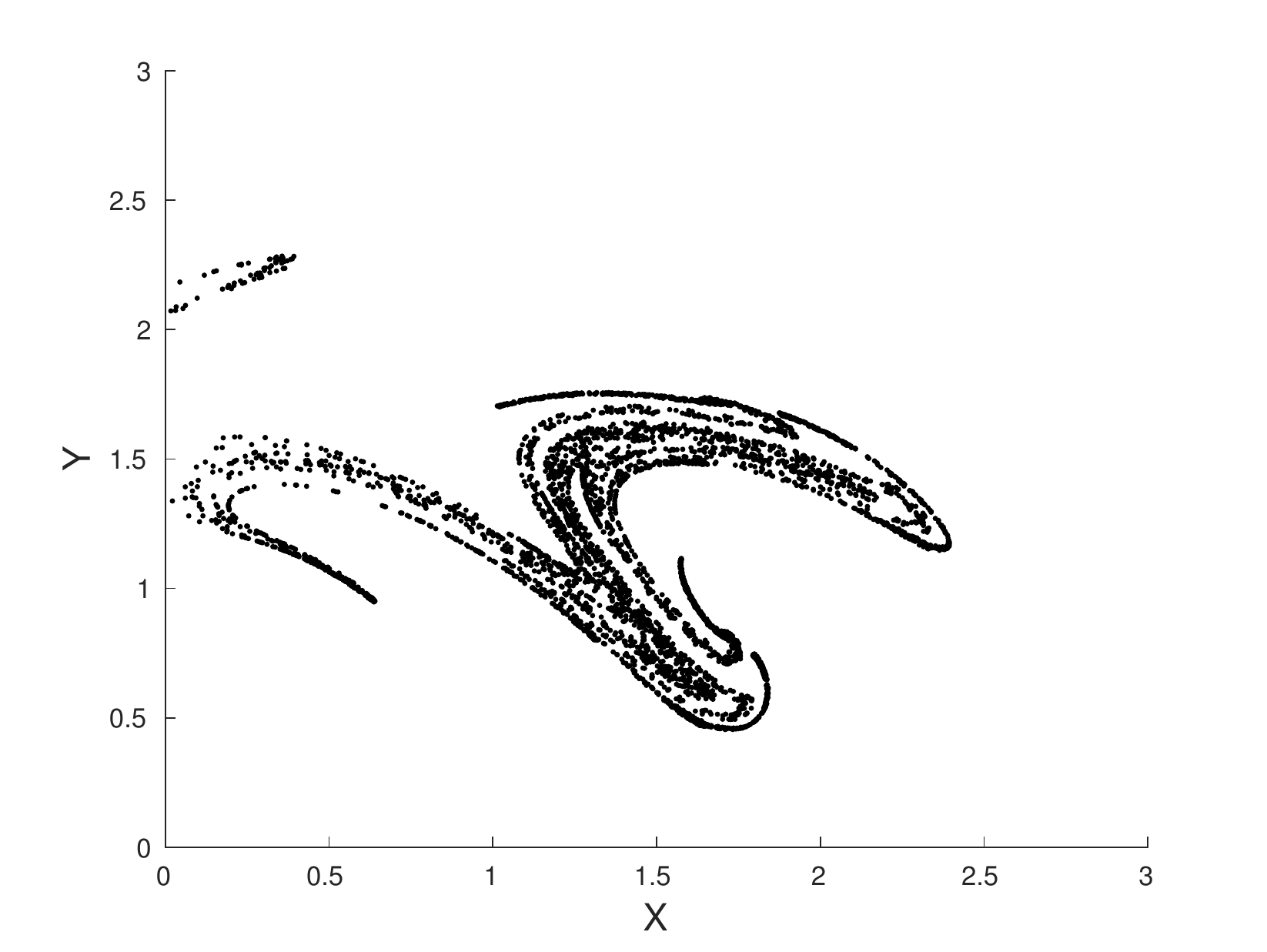}
  \caption{}
\end{subfigure}%
\begin{subfigure}{.5\textwidth}
  \includegraphics[width=\linewidth]{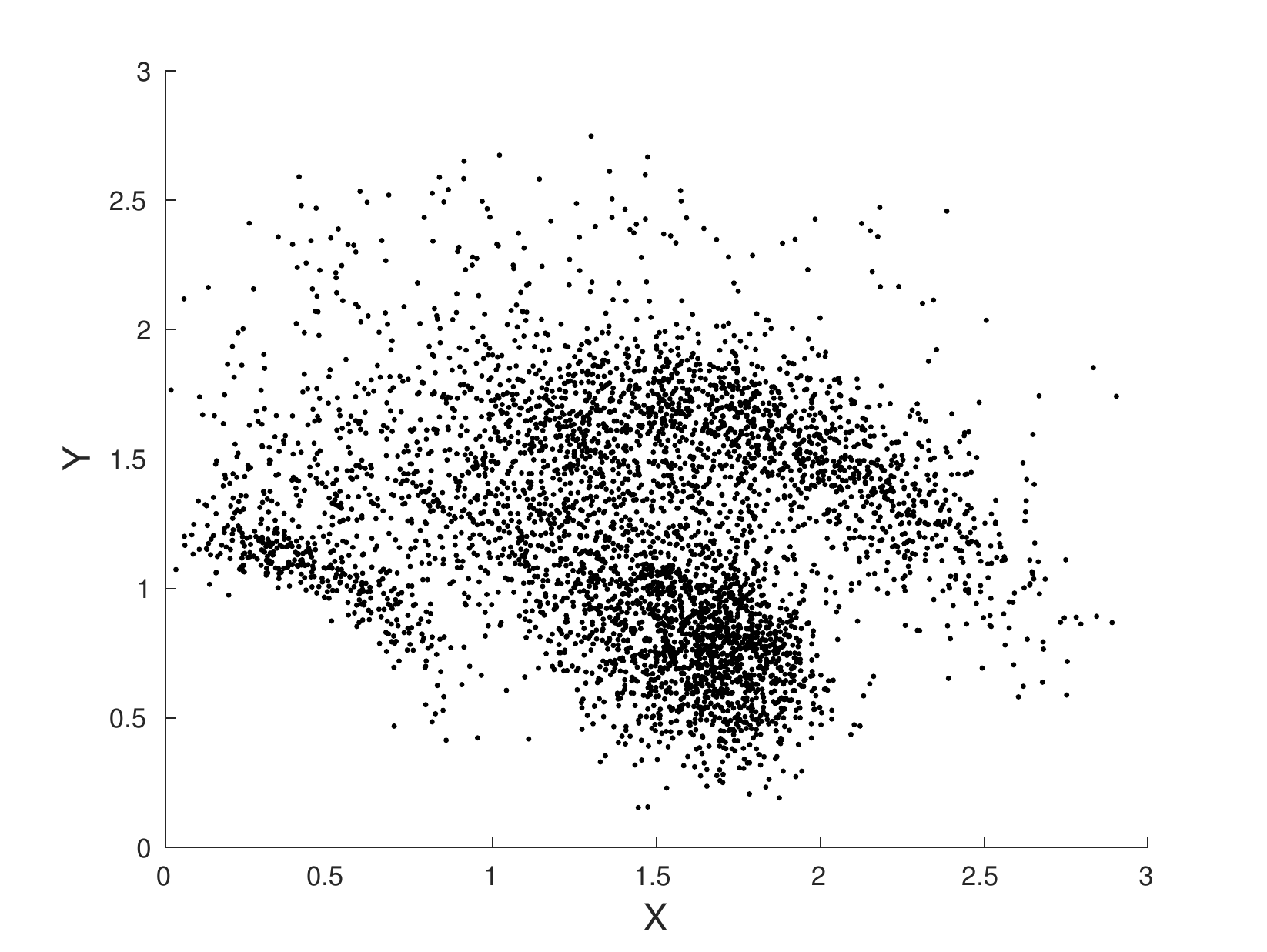}
  \caption{}
\end{subfigure}%
  \caption{Poincar\'e section in $Z=0$ of a) coupled model; b) uncoupled model; c) 1st order parametrization; d) 2nd order parametrization.}
  \label{PoinSecZ0}
\end{figure*}

We first provide a qualitative overview of the performance of the parametrization by investigating a few Poincar\'e sections, which provide a convenient and widely used method to visualize the dynamics of a system in a two-dimensional plot \citep{Eckmann1985,Ott1993}; typically, the plane chosen for the section of Lorenz 84 is $Z=0$. Fig.\ref{PoinSecZ0}a) shows the Poincar\'e section at $Z=0$ of the variables $X$, $Y$ of the coupled model given in Eqs.\ref{L84XC}-\ref{L63zC}. Panels b) of the same figure shows the Poincar\'e section of the Lorenz 84 model obtained by removing the coupling with the Lorenz 63 model. Finally, Panels c) and d) show the Poincar\'e sections of the modified Lorenz 84 models obtained by adding the first and second order parametrization, respectively. Visual inspection suggests that the second order parametrization does a good job in reproducing the properties of the full coupled model.

\begin{figure*}
\begin{subfigure}{.5\textwidth}
  \centering
  \includegraphics[width=\linewidth]{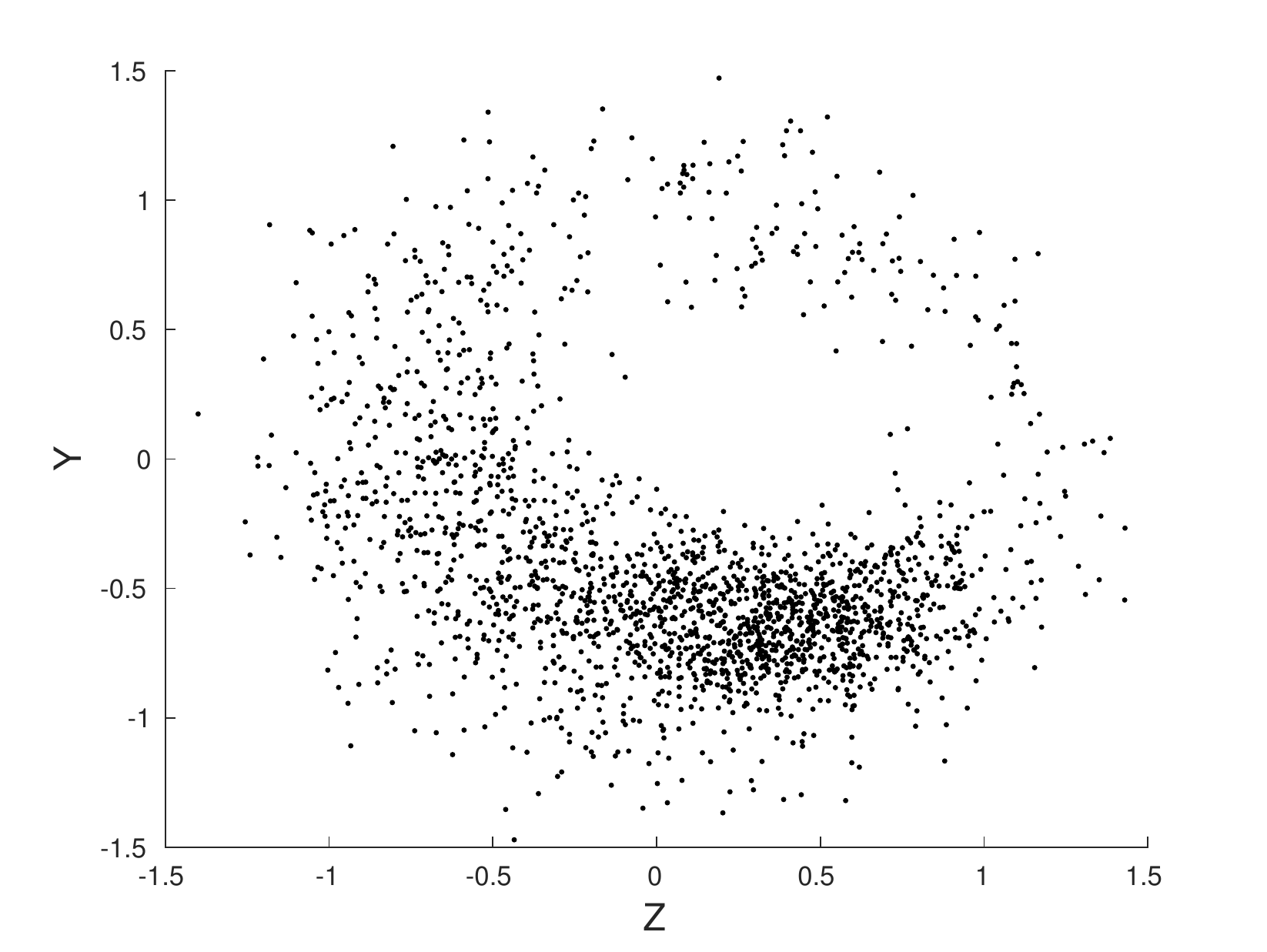}
  \caption{}
\end{subfigure}%
\begin{subfigure}{.5\textwidth}
  \includegraphics[width=\linewidth]{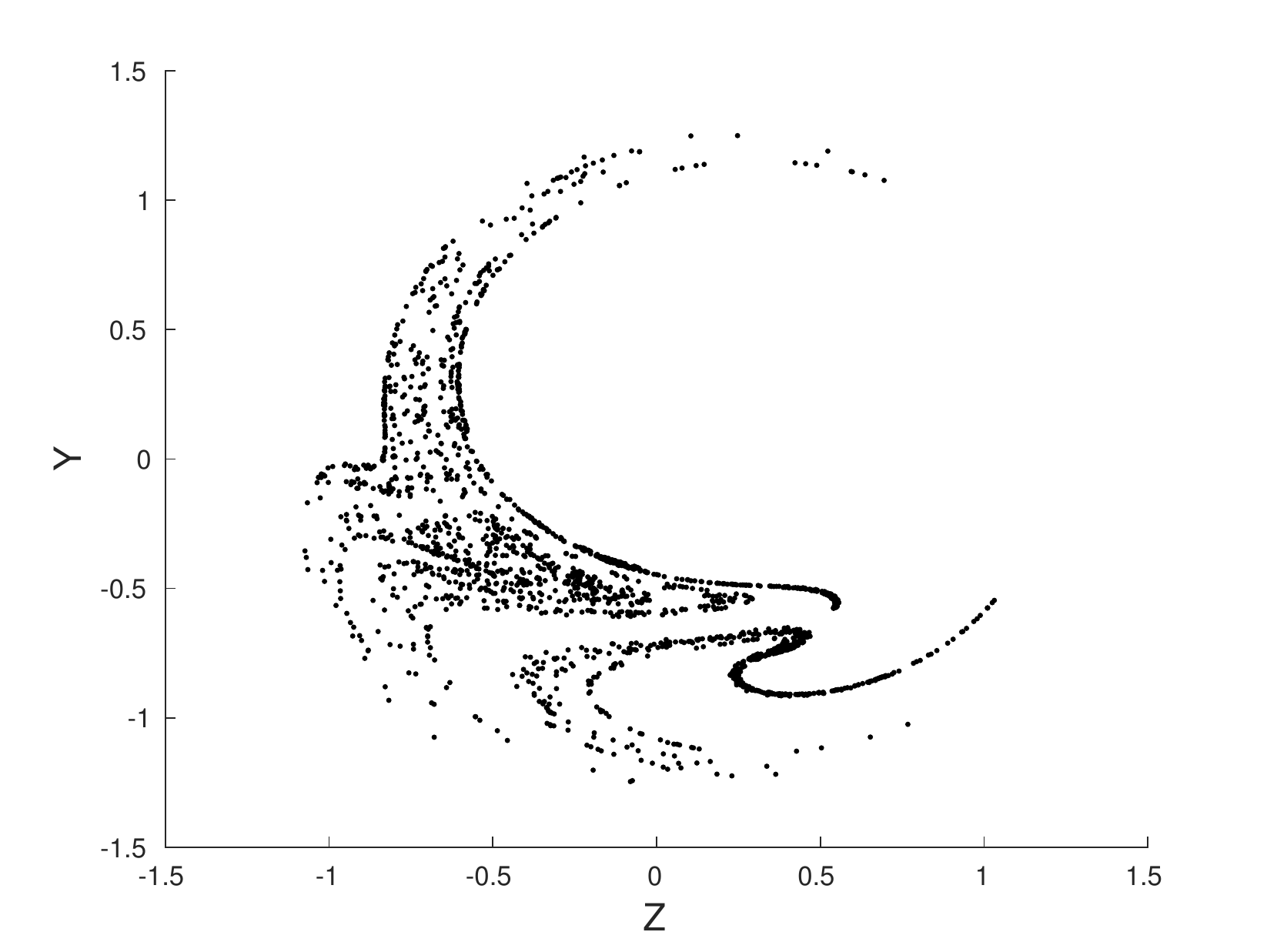}
  \caption{}
\end{subfigure}\\
\begin{subfigure}{.5\textwidth}
  \centering
  \includegraphics[width=\linewidth]{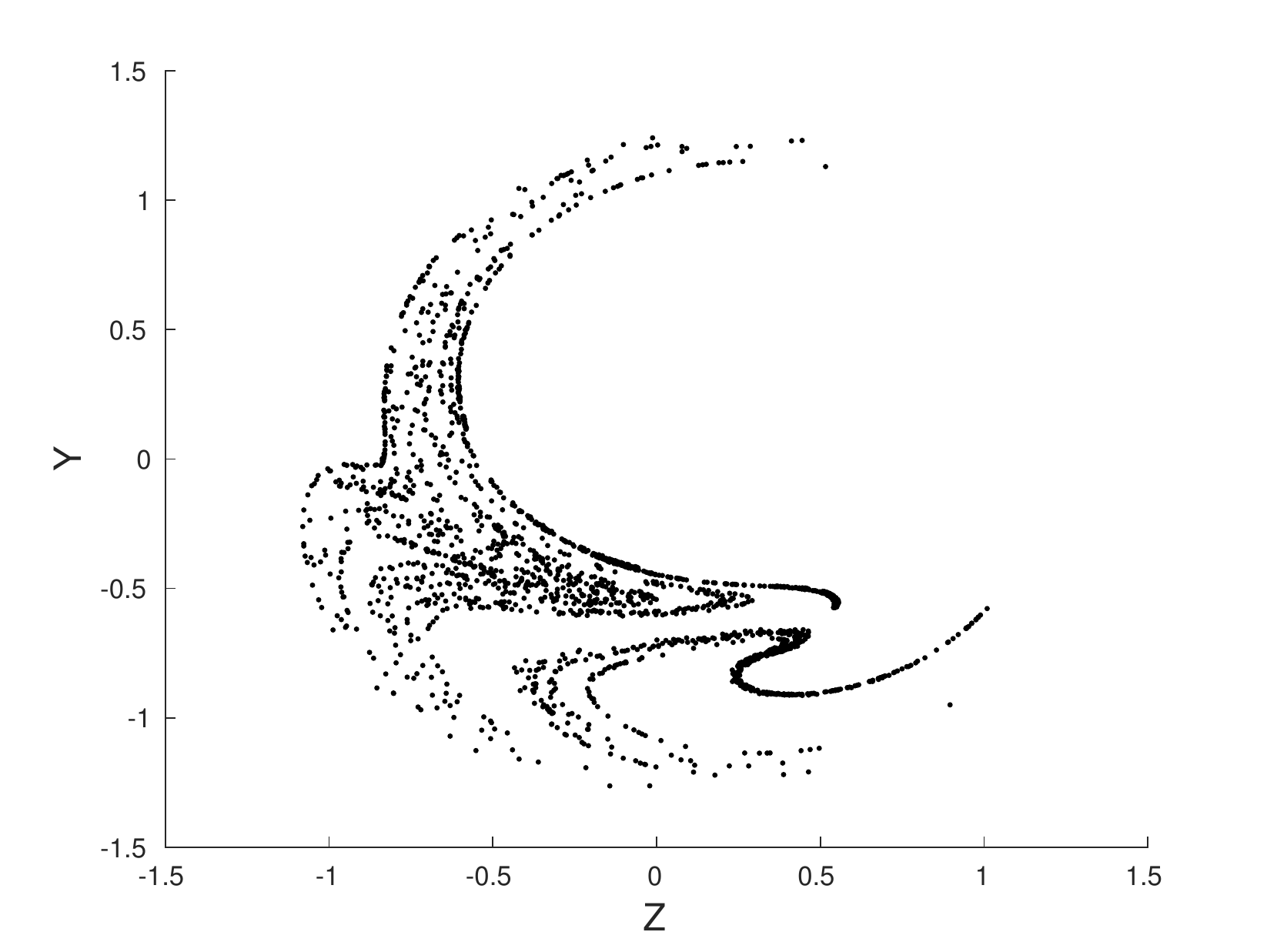}
  \caption{}
\end{subfigure}%
\begin{subfigure}{.5\textwidth}
  \includegraphics[width=\linewidth]{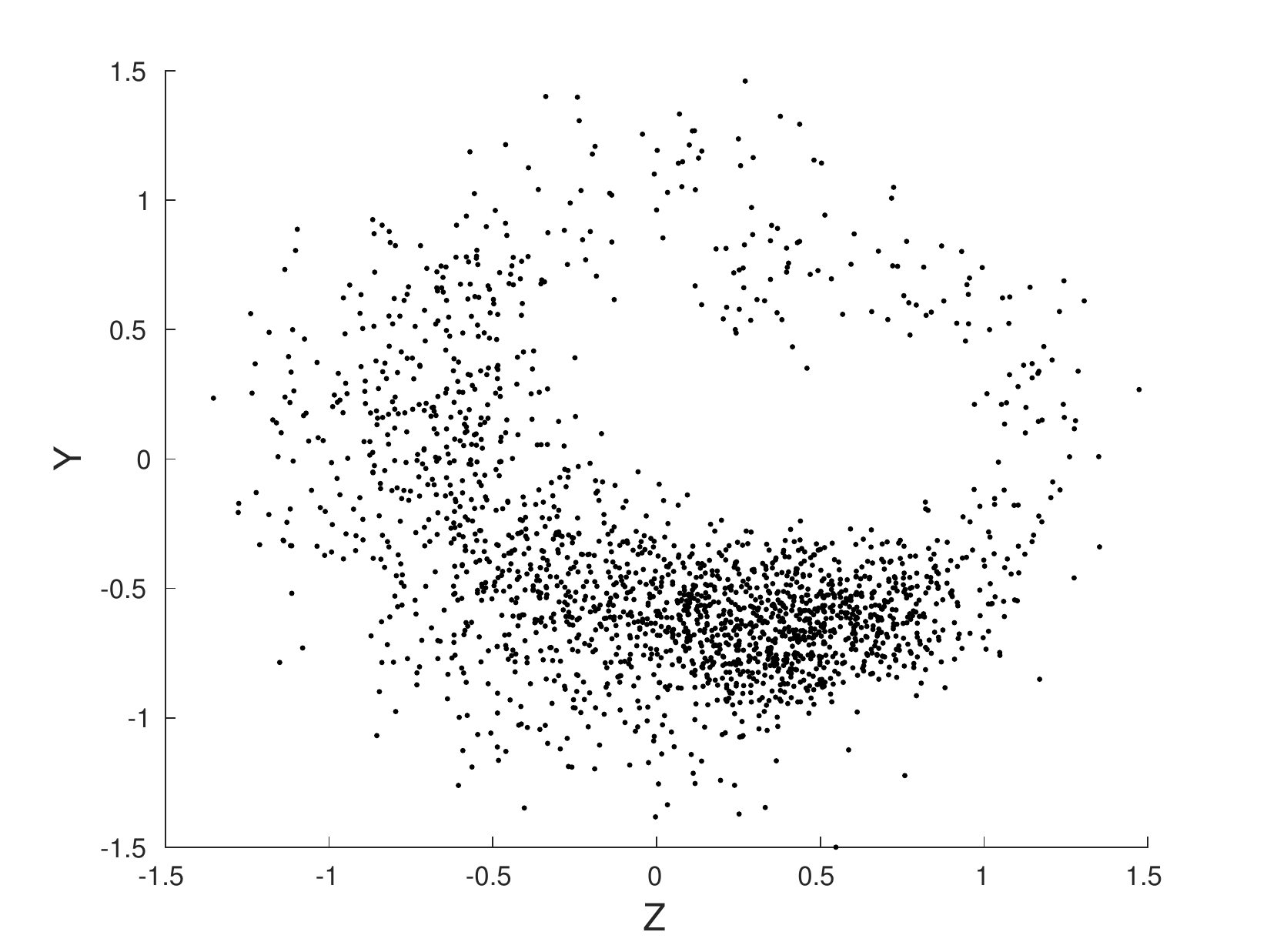}
  \caption{}
\end{subfigure}%
  \caption{Poincar\'e section in $X=1$ of a) coupled model; b) uncoupled model; c) 1st order parametrization; d) 2nd order parametrization.}
  \label{PoinSecX1}
\end{figure*}

Metaphorically, our parametrization aims at describing as accurately as possible the impact of "convection" on the "westerlies". It is insightful to look at how it affects the properties of the two variables - $X$ and $Y$ - that are not directly impacted by it. This amounts to looking at the impact of the parametrization of "convection" on the "large scale planetary waves" and, consequently, on the "large scale heat transport". Therefore, we look into $X=constant$ Poincar\'e section, in order to highlight the properties of $Y$ and $Z$. The four panels in Fig.\ref{PoinSecX1} are structured as in Fig.\ref{PoinSecZ0} and depict the Poncar\'e section of $X=1$. Also in this case the second order parametrization provides a far better match to the coupled model, featuring a remarkable ability in the reproducing the main features of the pattern of points.

\begin{figure*}
\begin{subfigure}{.5\textwidth}
  \centering
  \includegraphics[width=\linewidth]{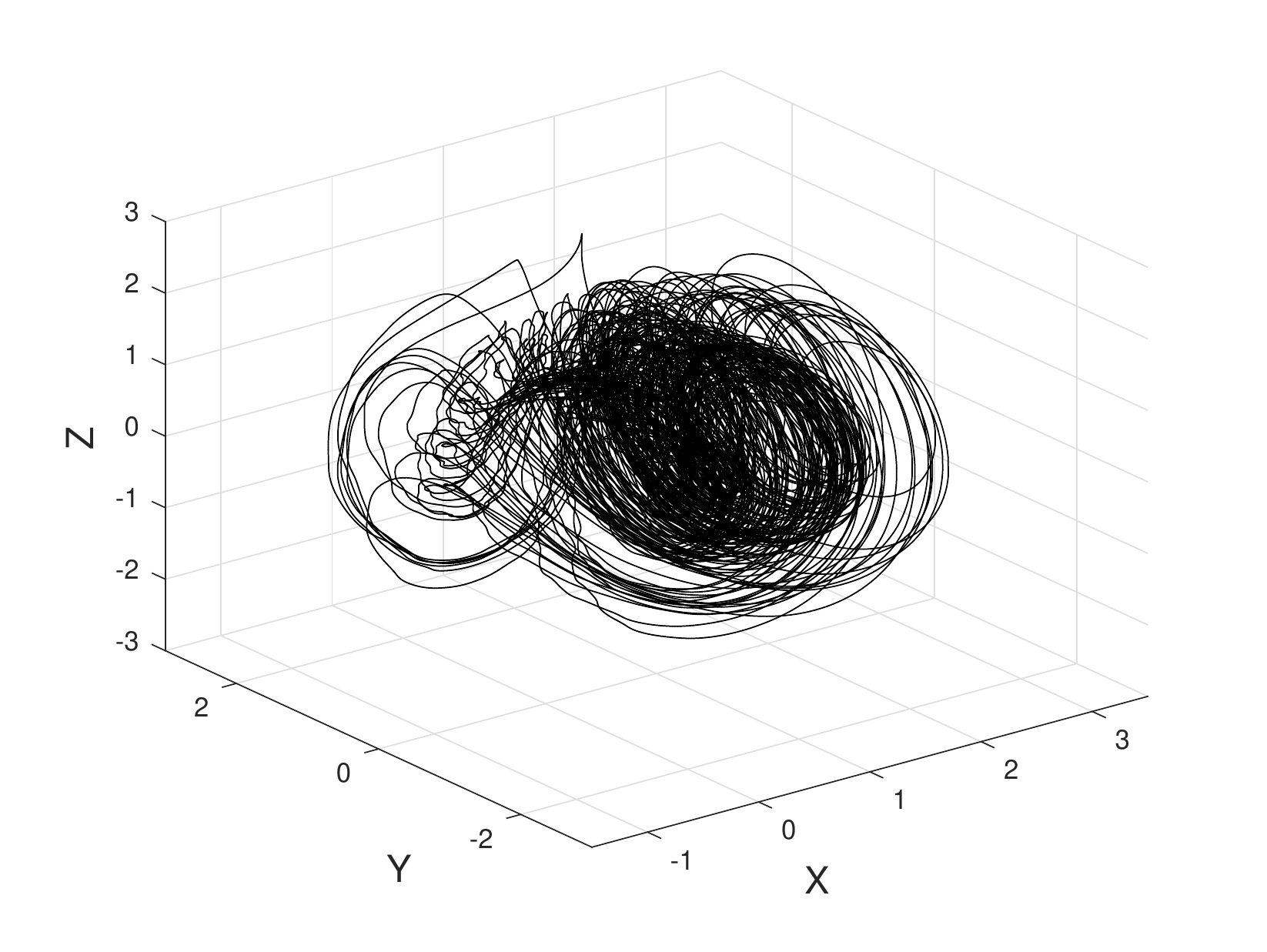}
  \caption{}
\end{subfigure}%
\begin{subfigure}{.5\textwidth}
  \includegraphics[width=\linewidth]{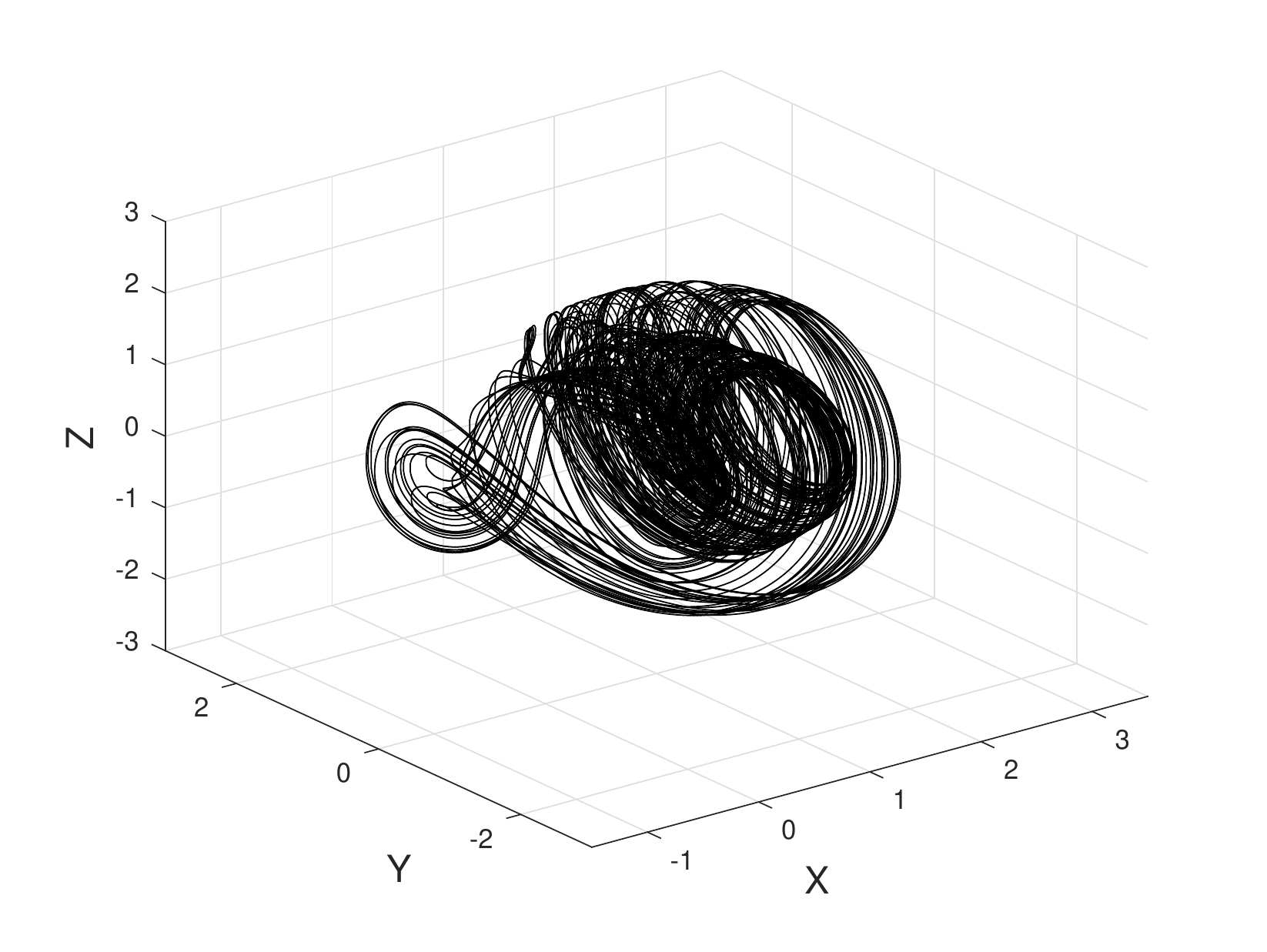}
  \caption{}
\end{subfigure}\\
\begin{subfigure}{.5\textwidth}
  \centering
  \includegraphics[width=\linewidth]{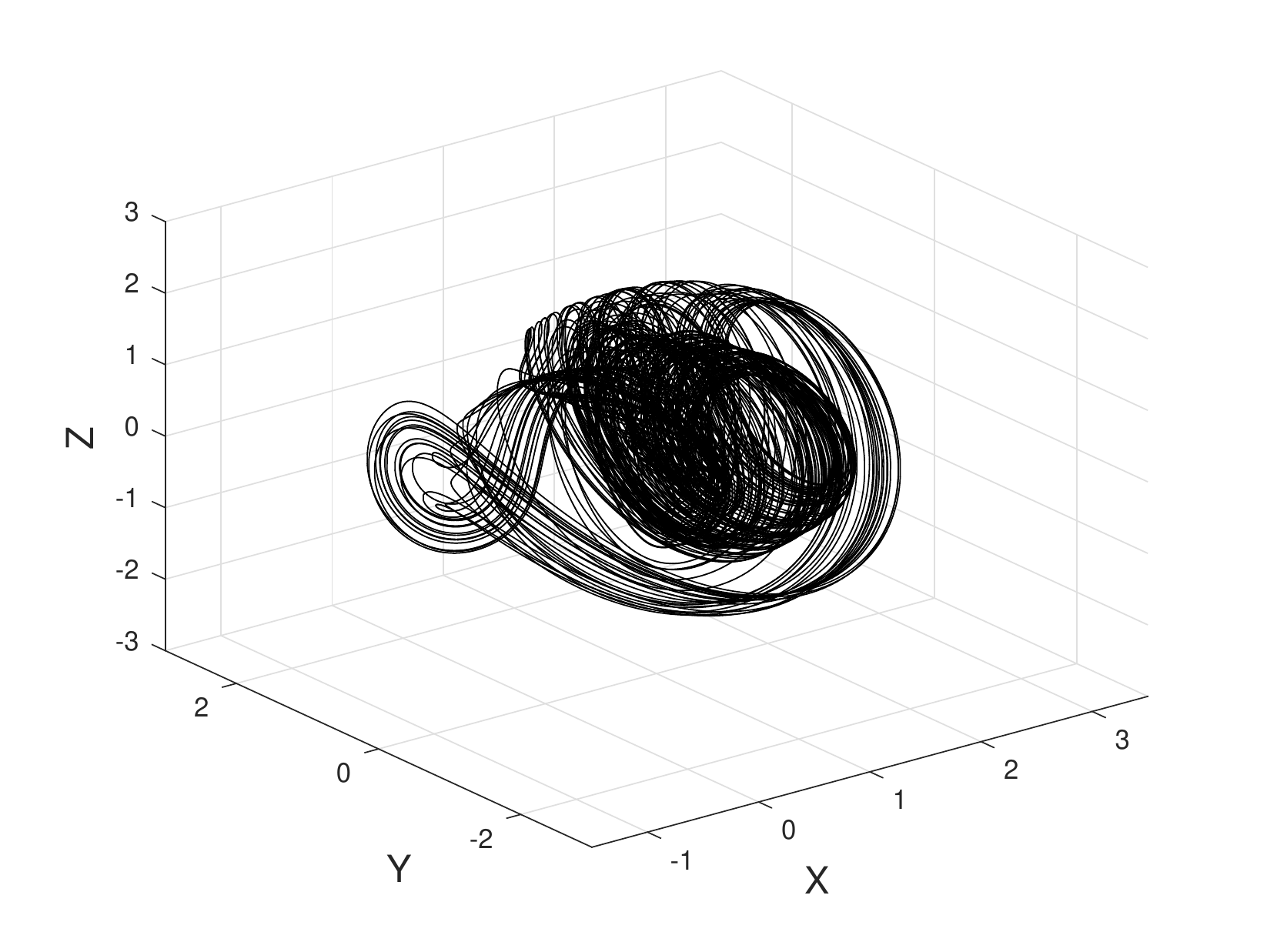}
  \caption{}
\end{subfigure}%
\begin{subfigure}{.5\textwidth}
  \includegraphics[width=\linewidth]{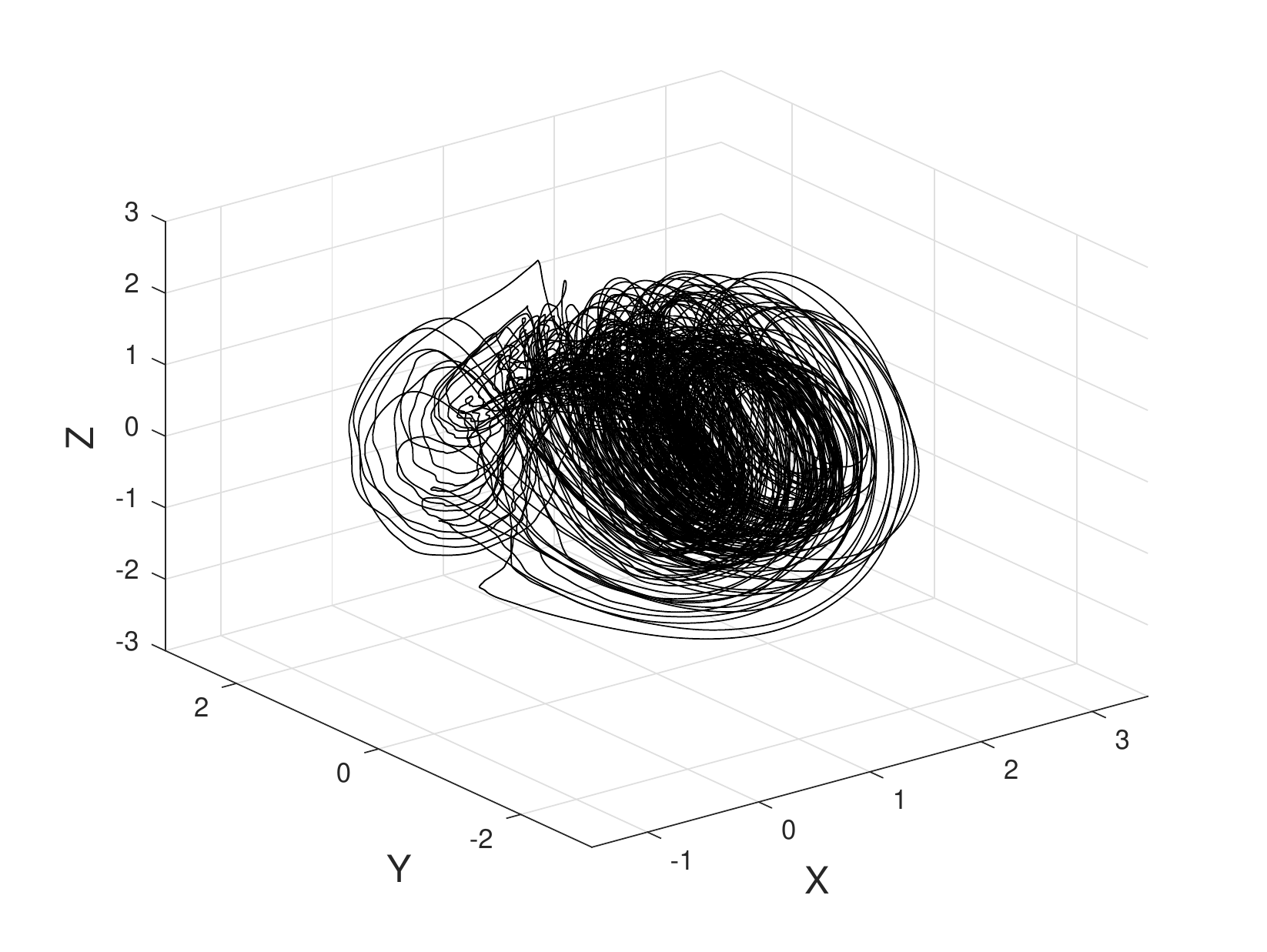}
  \caption{}
\end{subfigure}%
  \caption{$3D$ view of the attractor of a) coupled model; b) uncoupled model; c) 1st order parametrization; d) 2nd order parametrization.}
  \label{Attractor3D}
\end{figure*}

In order to provide further qualitative evidence of our results, in four panels of Fig.\ref{Attractor3D} we show the trajectories in the phase space of the $X$, $Y$, and $Z$ variables for the four considered models. For the sake of clarity, the plots are created using just $5$ years ($365$ time units). In the case of the coupled model the attractor covers a large portion of the phase space and the projections are less regular than for the uncoupled case. The second order parametrization successfully imitates this feature, while the simple deterministic correction, once again, is completely inadequate.

\subsection{Evaluation of the Performance of the Parametrization} \label{EvalPerfPar}
Further to the qualitative inspection, we provide here quantitative comparisons to support our study. All the remaining simulations in this section are run for $100$ years ($7300$ time units) with a time step equal to $0.005$; thus, each attractor is constructed with $1460000$ points.

We first look into the probability densities (PDFs) of the variables $X$, $Y$ and $Z$, which describe, loosely speaking, our climate.

\begin{figure*}
  \includegraphics[width=\linewidth]{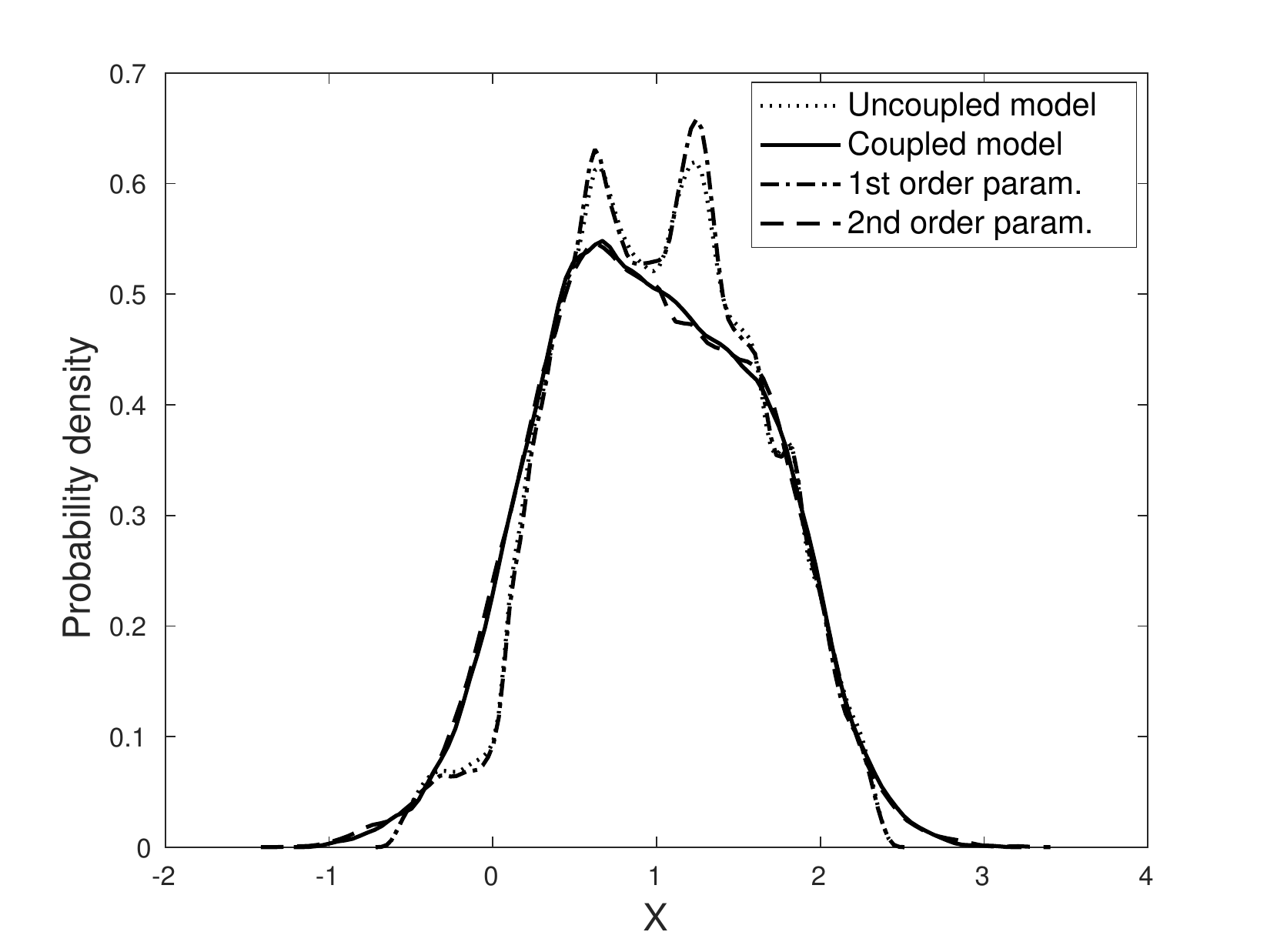}
  \caption{Probability density of the $X$ variable.}
  \label{ProbDensX}
\end{figure*}

\begin{figure*}
  \includegraphics[width=\linewidth]{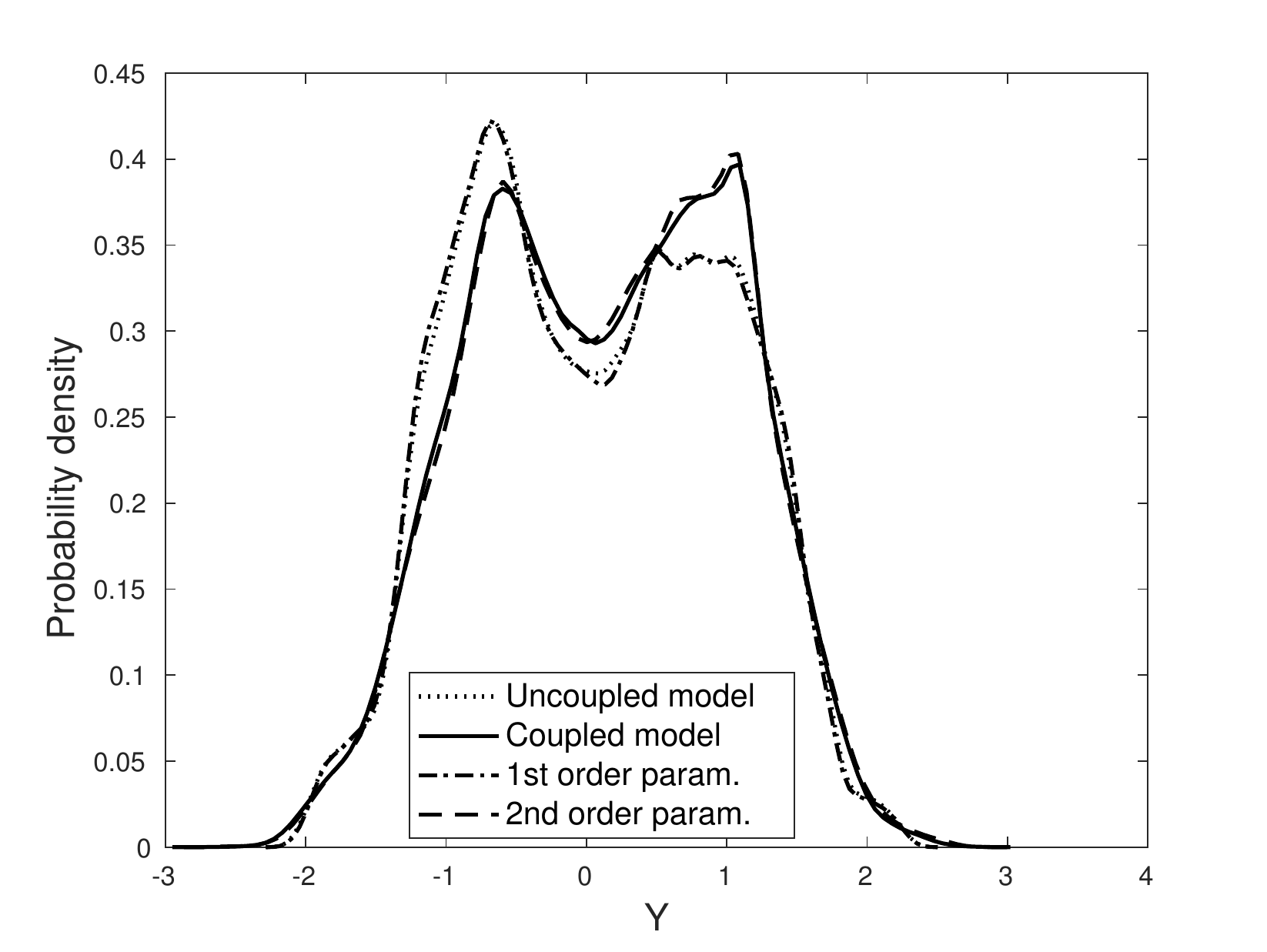}
  \caption{Probability density of the $Y$ variable.}
  \label{ProbDensY}
\end{figure*}

\begin{figure*}
  \includegraphics[width=\linewidth]{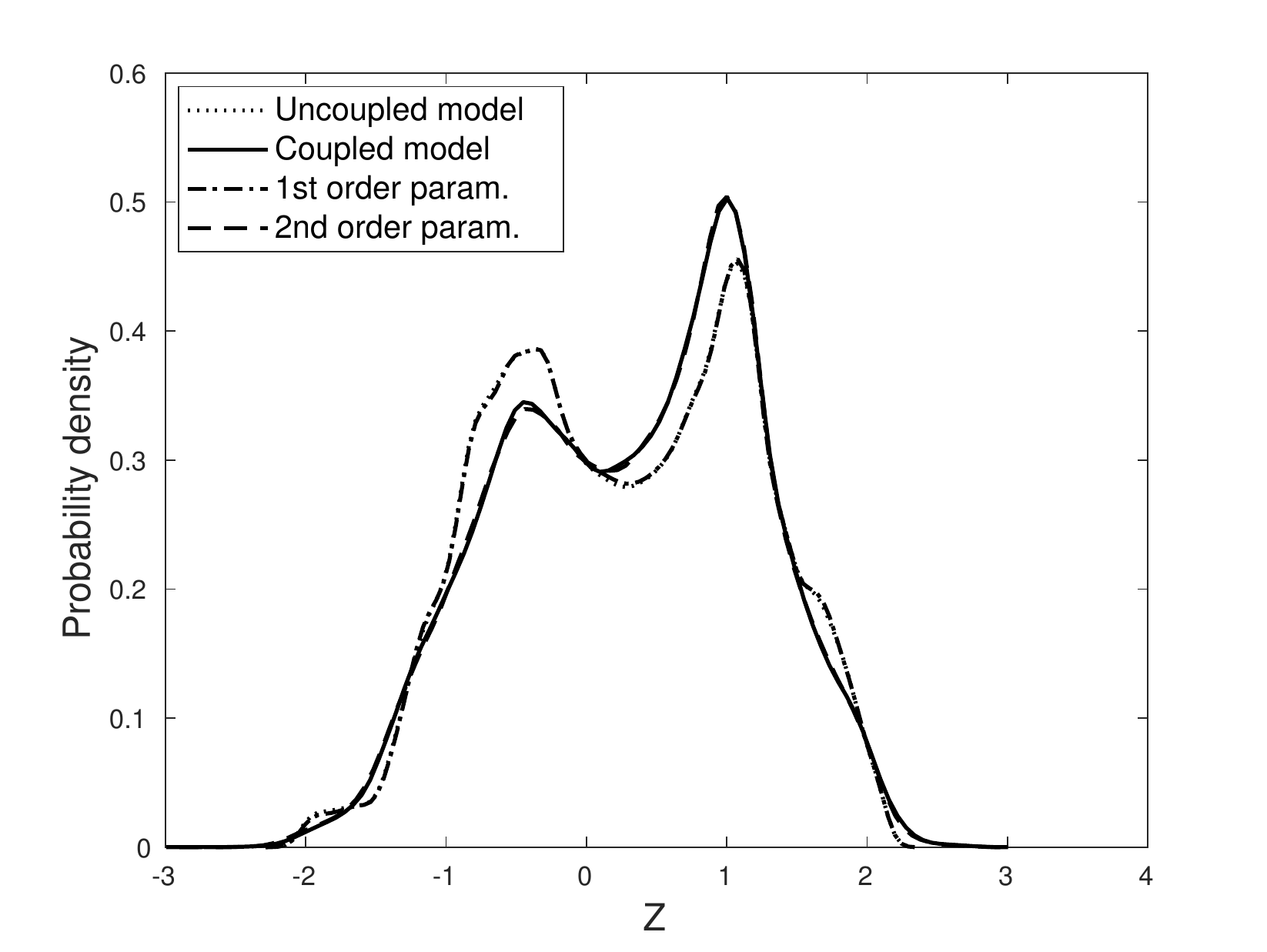}
  \caption{Probability density of the $Z$ variable.}
  \label{ProbDensZ}
\end{figure*}

Fig.\ref{ProbDensX} shows the PDF of the $X$ variable, for the four considered models. As expected, the second order parametrization allows for reconstructing with great accuracy the statistics of the coupled model. The bimodality of the uncoupled Lorenz 84 model is reproduced by the model featuring the first order parametrization, while the second order model predicts accurately the unimodal distribution shown by the coupled model. The PDFs for $Y$ and $Z$ variables are shown in Figs.\ref{ProbDensY}-\ref{ProbDensZ}, respectively. Also here, where the external forcing does not destroy the bimodality of the distributions found in the uncoupled case, WL parametrization leads to a very good approximation of the properties of the coupled model. In particular, the tails of the distributions are replicated with a high level of precision, making possible to seemingly reproduce with good accuracy the extreme values of the variables. This is a matter worth investigating in a separate study. Note that, since the WL parametrization is constructed to have skill for all observables, it is not so surprising that it can perform well also far away from the bulk of the statistics.

Aside from the analysis of the PDF, a further statistical investigation can be provided by looking into the numerical results provided by first moments of the variables and their uncertainty, which is the standard deviation derived from the analysis of an ensemble of runs. We have performed just ten runs, but our results are very robust. The results for the statistics of the first two moments are reported in Table \ref{tab:Xvalues}: all the quantities inspected clearly show that the second order parametrization allows for reproducing very accurately the moments statistics of the coupled model.

\begin{table}
  \caption{Expectation values for averages, variances and covariances with the correspondent standard deviations $\sigma$ (order of magnitude $-2$).}
  \label{tab:Xvalues}
\noindent\makebox[\textwidth]{

  \begin{tabular}{|l||c|c|c|c|}
    \hline
    Observables & Uncoupled model & 1st order parametrization & 2nd order parametrization & Coupled model\\
    \hline
    $\overline{X}\pm\sigma_{\overline{X}}$ & $101.5\pm0.4$ & $101.3\pm0.5$ & $97.2\pm0.3$ & $97.1\pm0.3$\\
    $\overline{Y}\pm\sigma_{\overline{Y}}$ & $6.1\pm0.8$ & $6.5\pm1.2$ & $13.7\pm0.7$ & $13.9\pm0.4$\\
    $\overline{Z}\pm\sigma_{\overline{Z}}$ & $27.0\pm0.2$ & $26.9\pm0.3$ & $31.0\pm0.2$ & $31.3\pm0.5$\\
    $var(X)\pm\sigma_{var(X)}$ & $34.9\pm0.8$ & $35.2\pm1.0$ & $43.6\pm0.7$ & $43.5\pm0.3$\\
    $var(Y)\pm\sigma_{var(Y)}$ & $84.4\pm0.1$ & $84.4\pm0.1$ & $82.8\pm0.4$ & $82.6\pm0.3$\\
    $var(Z)\pm\sigma_{var(Z)}$ & $82.6\pm0.1$ & $82.6\pm0.2$ & $81.5\pm0.3$ & $81.4\pm0.3$\\
    $cov(XY)\pm\sigma_{cov(XY)}$ & $-5.4\pm0.8$ & $-5.7\pm1.1$ & $-11.1\pm0.6$ & $-11.2\pm0.3$\\
    $cov(XZ)\pm\sigma_{cov(XZ)}$ & $-3.7\pm0.1$ & $-3.4\pm0.2$ & $-8.0\pm0.2$ & $-8.3\pm0.4$\\
    $cov(YZ)\pm\sigma_{cov(YZ)}$ & $-7.7\pm0.2$ & $-7.7\pm0.4$ & $-1.6\pm0.4$ & $-1.3\pm0.2$\\
    \hline
  \end{tabular}} 
\end{table}

\begin{figure*}
\begin{subfigure}{.5\textwidth}
  \centering
  \includegraphics[width=\linewidth]{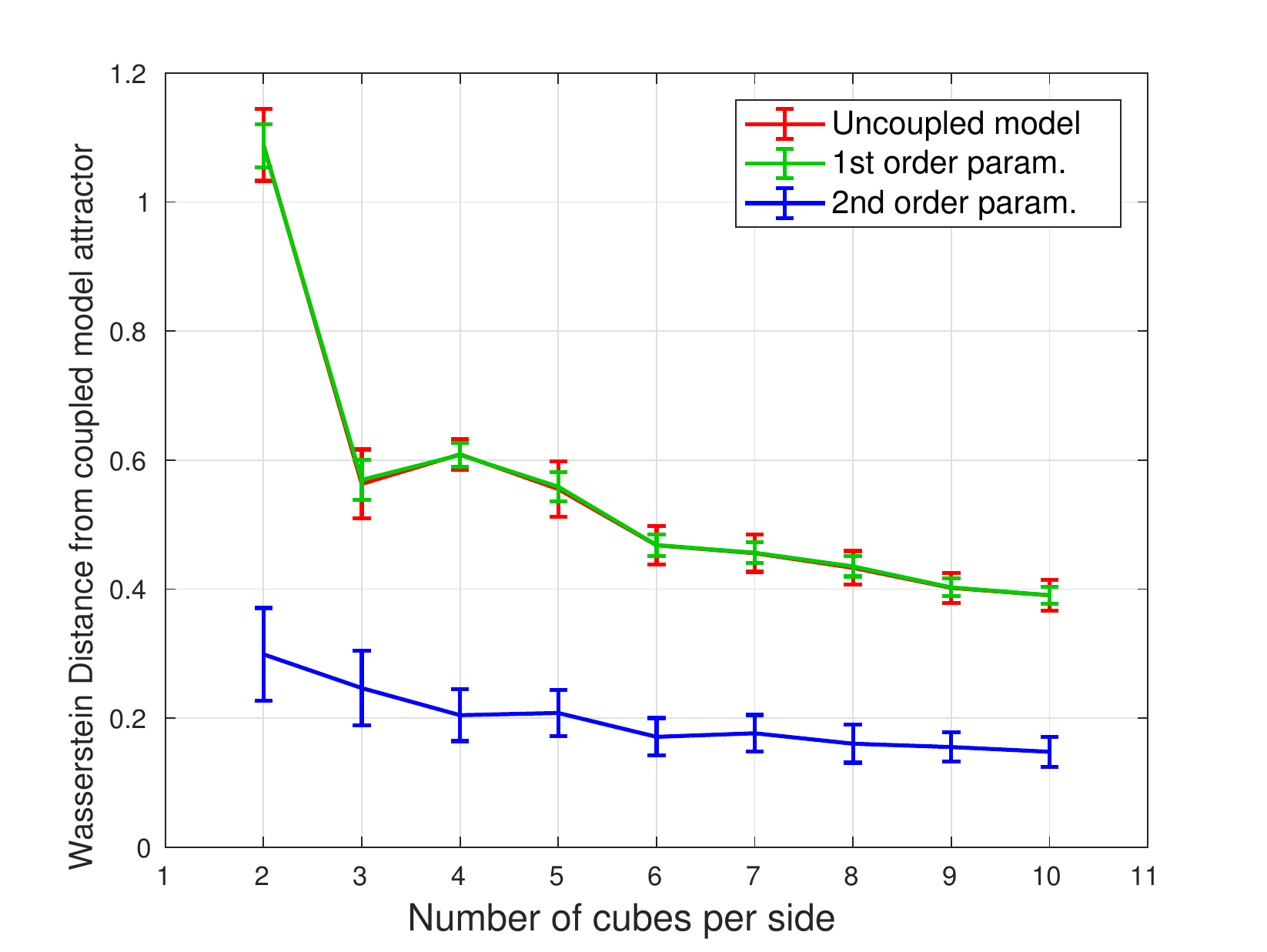}
  \caption{}
\end{subfigure}%
\begin{subfigure}{.5\textwidth}
  \includegraphics[width=\linewidth]{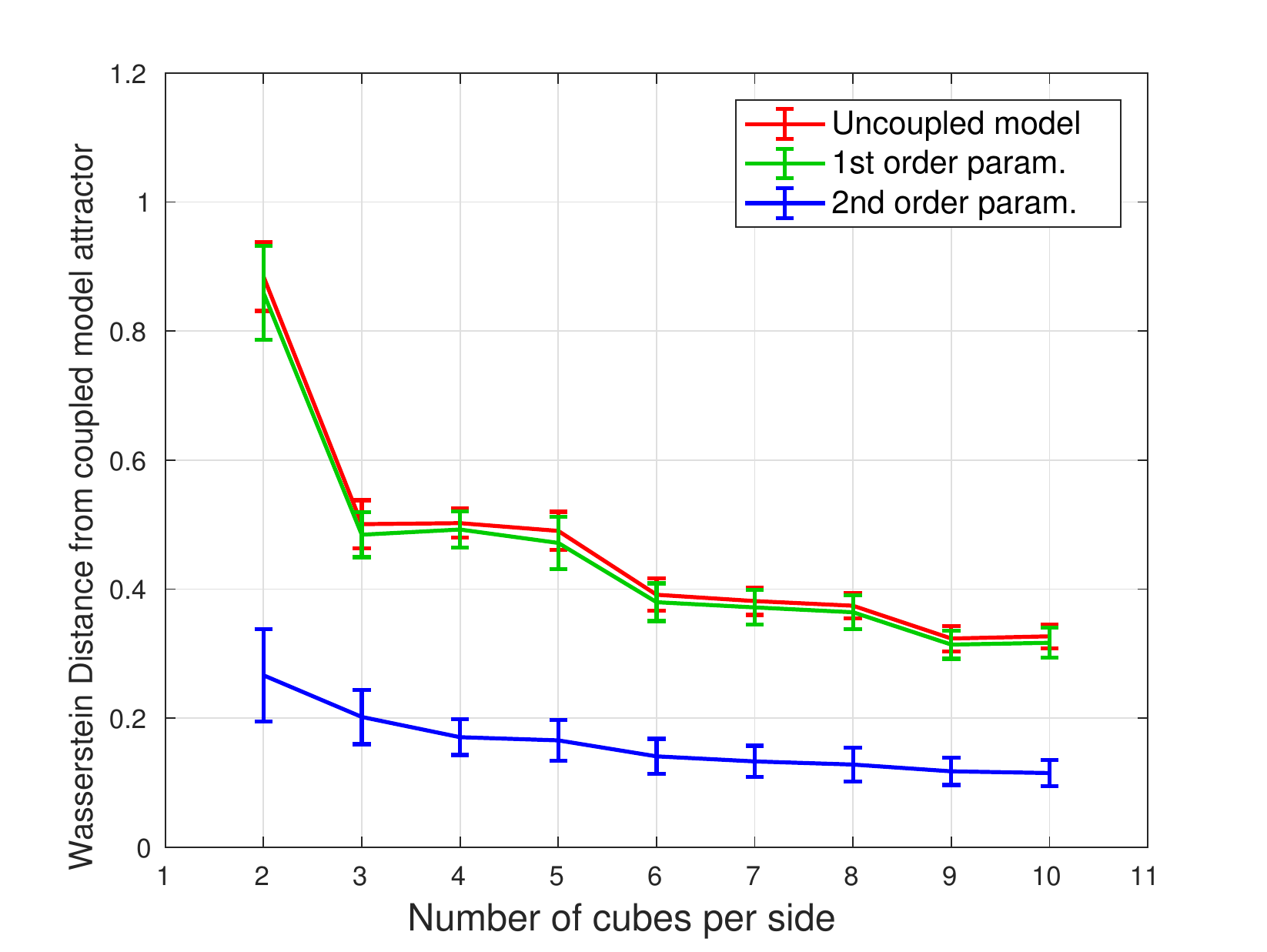}
  \caption{}
\end{subfigure}\\
\begin{subfigure}{.5\textwidth}
  \centering
  \includegraphics[width=\linewidth]{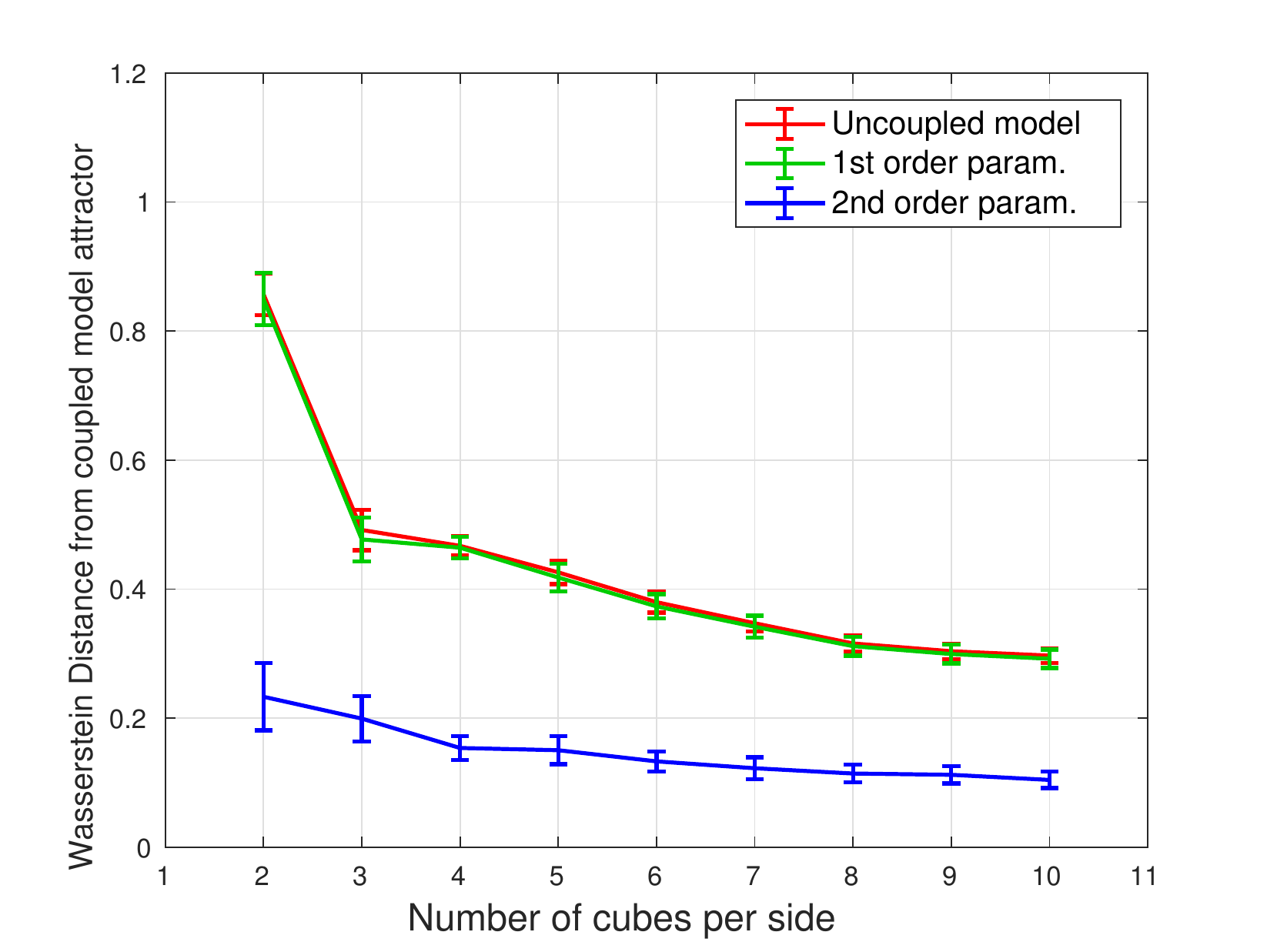}
  \caption{}
\end{subfigure}%
\begin{subfigure}{.5\textwidth}
  \includegraphics[width=\linewidth]{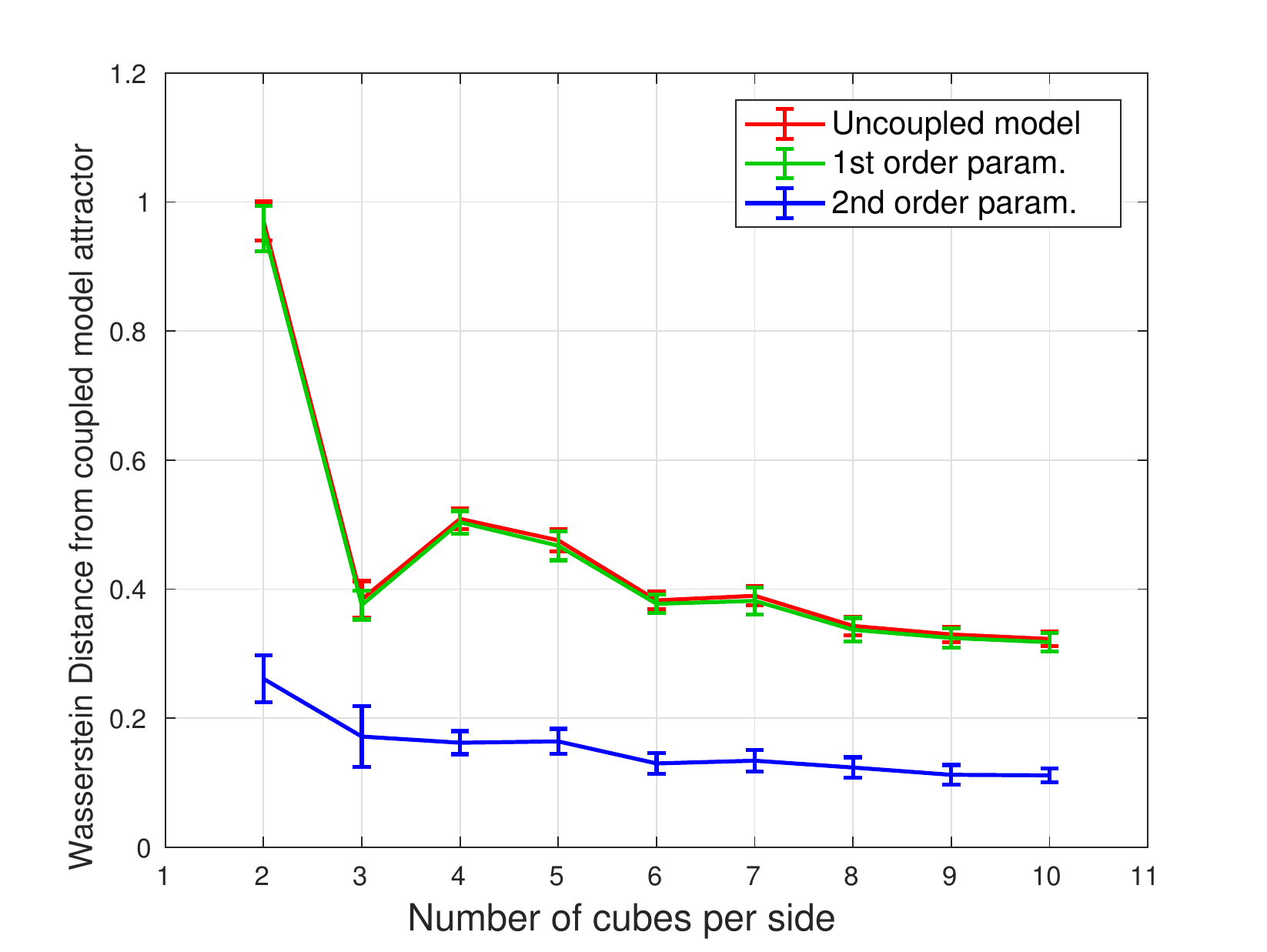}
  \caption{}
\end{subfigure}%
  \caption{Wasserstein distances from the coupled model with respect to number of cubes per side: a) $3D$ case; b) Projection on $XY$ plane; c) Projection on $XZ$ plane; d) Projection on $YZ$ plane.}
  \label{WD_n}
\end{figure*}

If the studied PDFs depart strongly from uni-modality, the analysis of the first moments can be of little utility, and it becomes hard to have a thorough understanding of the statistics adopting this point of view. As discussed above, we wish to supplement this simple analysis with a more robust evaluation of the performance of the parametrizations by taking into account the Wasserstein distance. A first issue to deal with in order compute the Wasserstein distance consists in carefully choosing the number of cubes used for the Ulam projection. Fig.\ref{WD_n}a shows the Wasserstein distance between the invariant measure of the coupled model projected on the $XYZ$ space and the invariant measure of the uncoupled Lorenz and of the models obtained using the first and second order parametrization. We find that for all choices of the coarse-graining the measure of the model with the second order parametrization is, by far, the closest one to the coupled case. Instead, the deterministic parametrization provides a negligible improvement with respect to the trivial case of considering the uncoupled model. What shown here gives a quantitative evaluation of the improved performance resulting from adding a stochastic parametrization. The second piece of information is that the distance has only a weak dependence on the degree of the coarse-graining and seems to approach its asymptotic value when finer resolutions of the Ulam projection are considered. This is encouraging as the findings one can obtain at low resolution seem to be already very meaningful and useful.

A well-known problem of Ulam's method is the fact that it can hardly be adapted to high dimensional spaces - this is the so-called curse of dimensionality. Additionally, evaluating the Wasserstein distance in high dimensions becomes itself computationally extremely challenging. In order to partially address these problems we repeat the analysis shown in Fig.\ref{WD_n}a) for the measures projected on the $XY$, $XZ$ and $YZ$ planes, thus constructing the so-called sliced Wasserstein distances. Results are reported in panels b), c), and d) of Fig.\ref{WD_n}, respectively. We find that, unsurprisingly, the distance of the projected measure is strictly lower than the distance in the full phase space, \textit{ceteris paribus}. What is more interesting is that all the observations we made for Fig.\ref{WD_n}a) apply for the other panels. Therefore, it seems reasonable to argue that studying the Wasserstein distance for projected spaces might provide useful information also on the full system.

In order to extend the scope of our study we have repeated the analysis described above for the case $\tau=\frac{1}{6}$. Such a choice implies that the model responsible for the forcing has a internal time scale which is comparable to the one of the model of interest. We remark that the WL parametrization, as discussed in \citep{Vissio2018}, is not based on any assumption of time scale separation between the variables of interest and the variables we want to parametrize. We report below only the main results for the sake of conciseness.

\begin{figure*}
\begin{subfigure}{.5\textwidth}
  \centering
  \includegraphics[width=\linewidth]{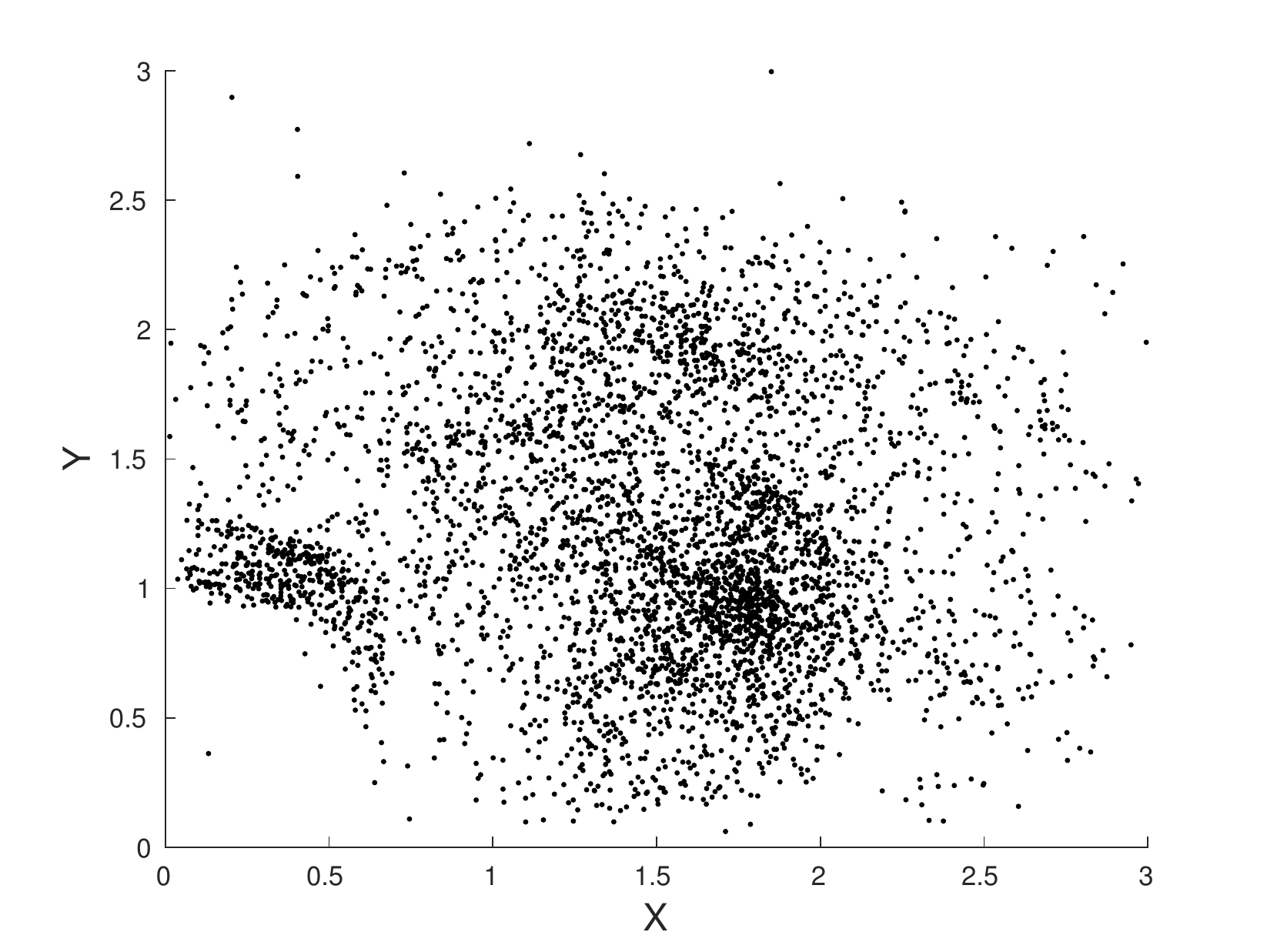}
  \caption{}
\end{subfigure}%
\begin{subfigure}{.5\textwidth}
  \includegraphics[width=\linewidth]{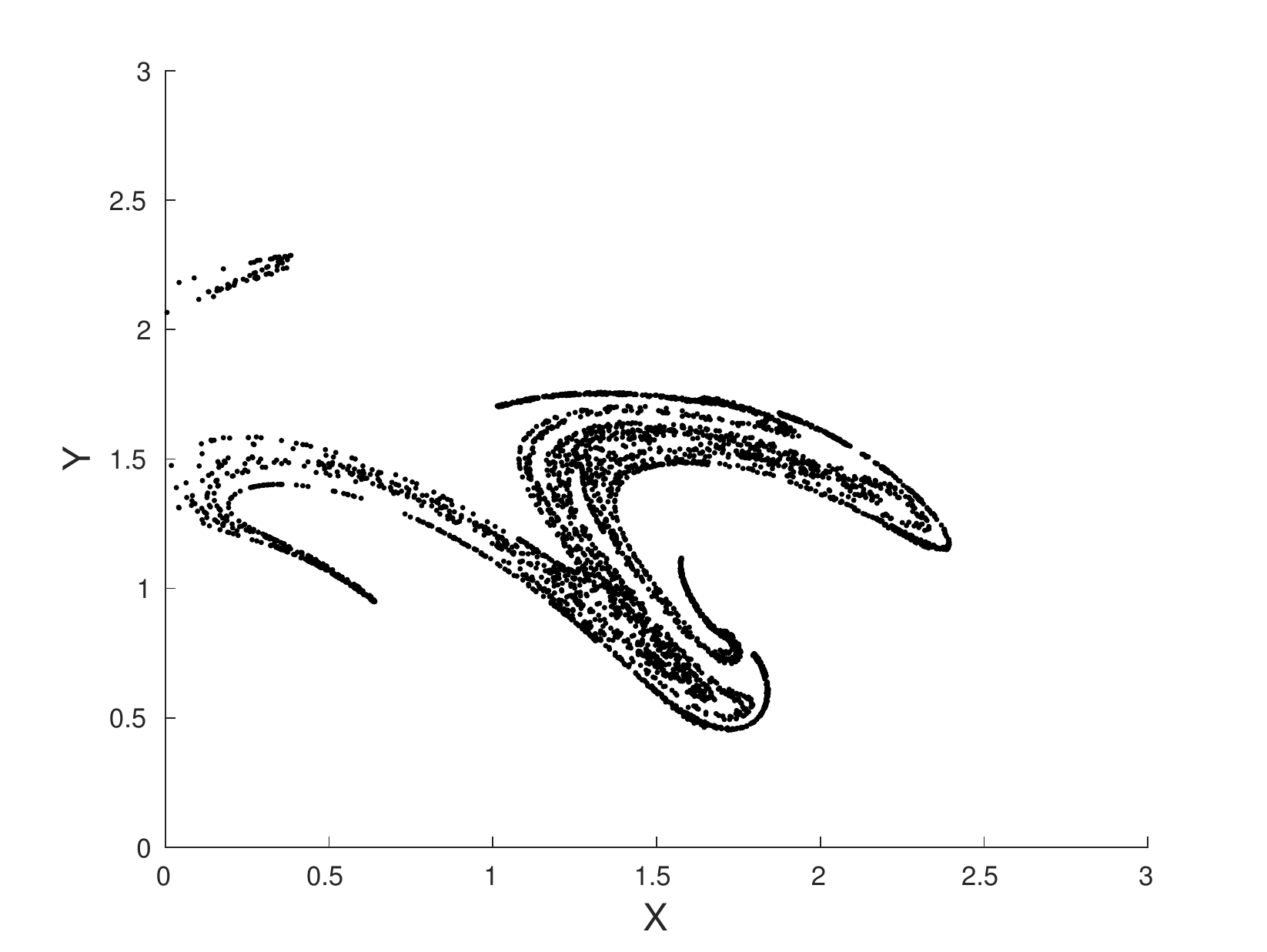}
  \caption{}
\end{subfigure}\\
\begin{subfigure}{.5\textwidth}
  \centering
  \includegraphics[width=\linewidth]{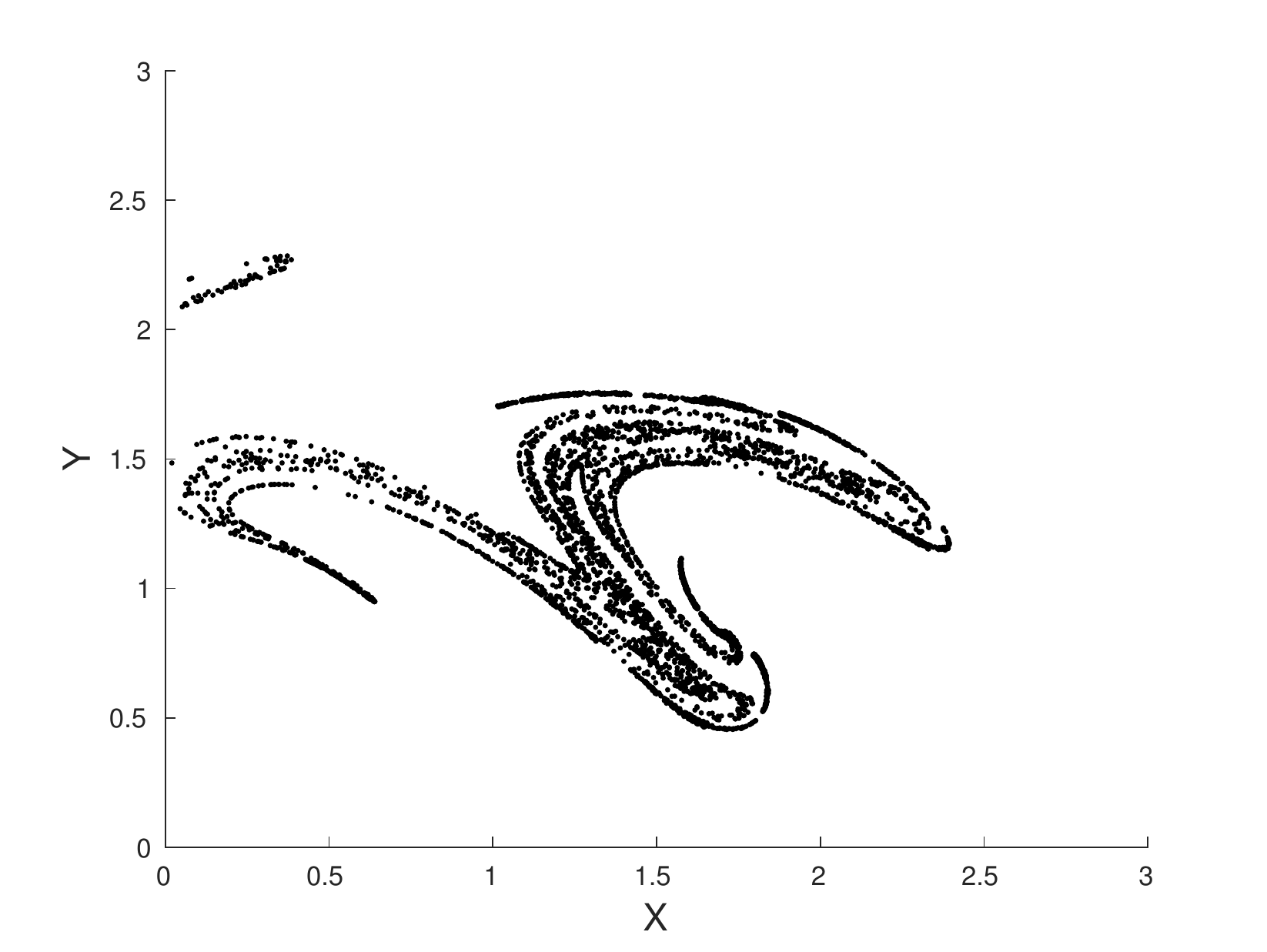}
  \caption{}
\end{subfigure}%
\begin{subfigure}{.5\textwidth}
  \includegraphics[width=\linewidth]{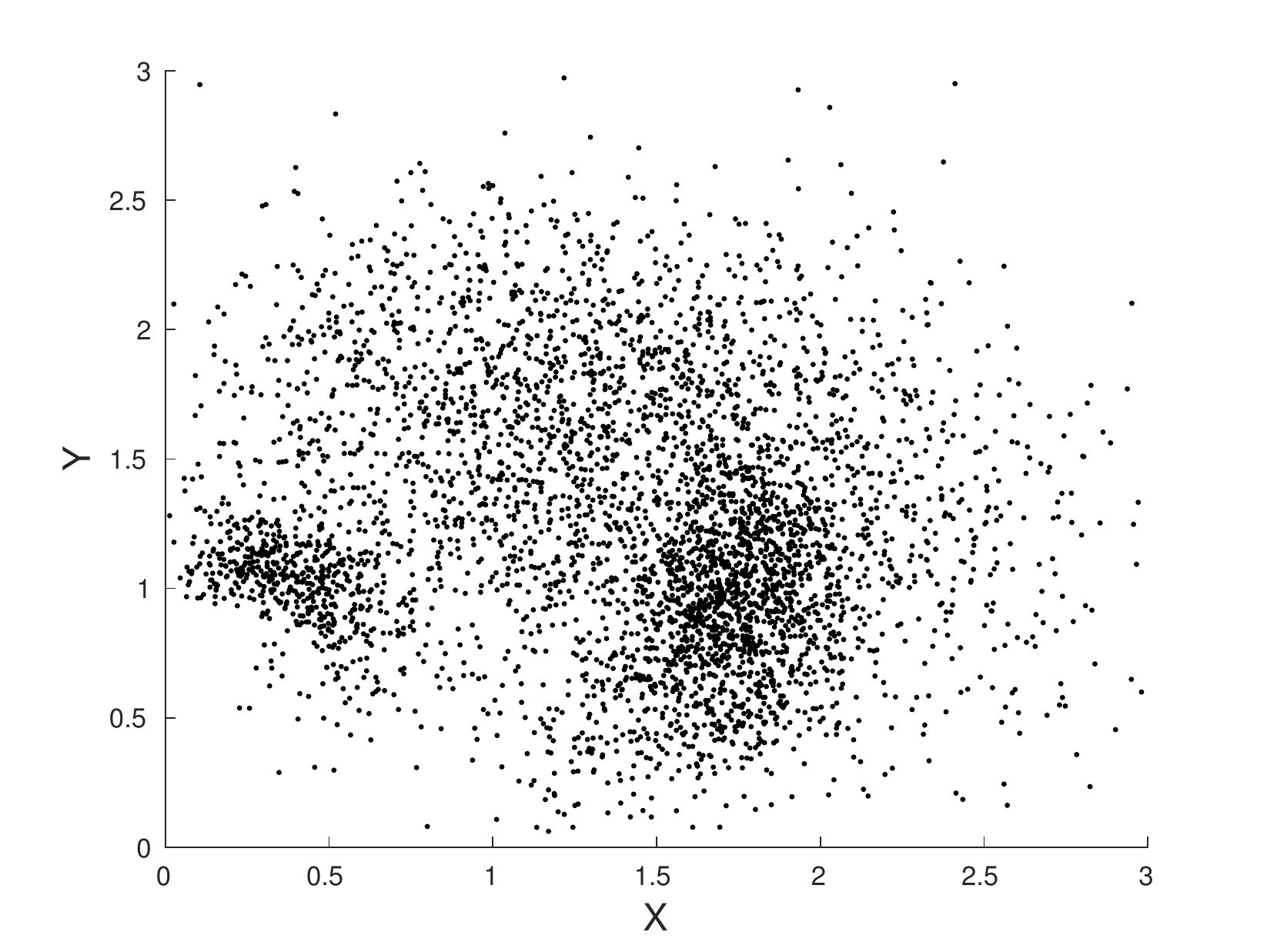}
  \caption{}
\end{subfigure}%
  \caption{Poincar\'e section in $Z=0$ of a) coupled model; b) uncoupled model; c) 1st order parametrization; d) 2nd order parametrization.}
  \label{PoinSecZ0tau05}
  
\end{figure*}

Figures \ref{PoinSecZ0tau05}a)-d) show the Poincar\'e sections in $Z=0$ for all the considered models. In the case of the coupled system, most of the fine structure one finds in the uncoupled model is lost, and we basically have a cloud of points with weaker features than what shown in Figure  \ref{PoinSecZ0} for $\tau=5$. Nonetheless, also in this case the model with the second order parametrization  reproduces (visually) quite well what shown in Panel a), and, in particular, shows matching regions where the density of the points is higher.



\begin{figure*}
\begin{subfigure}{.5\textwidth}
  \centering
  \includegraphics[width=\linewidth]{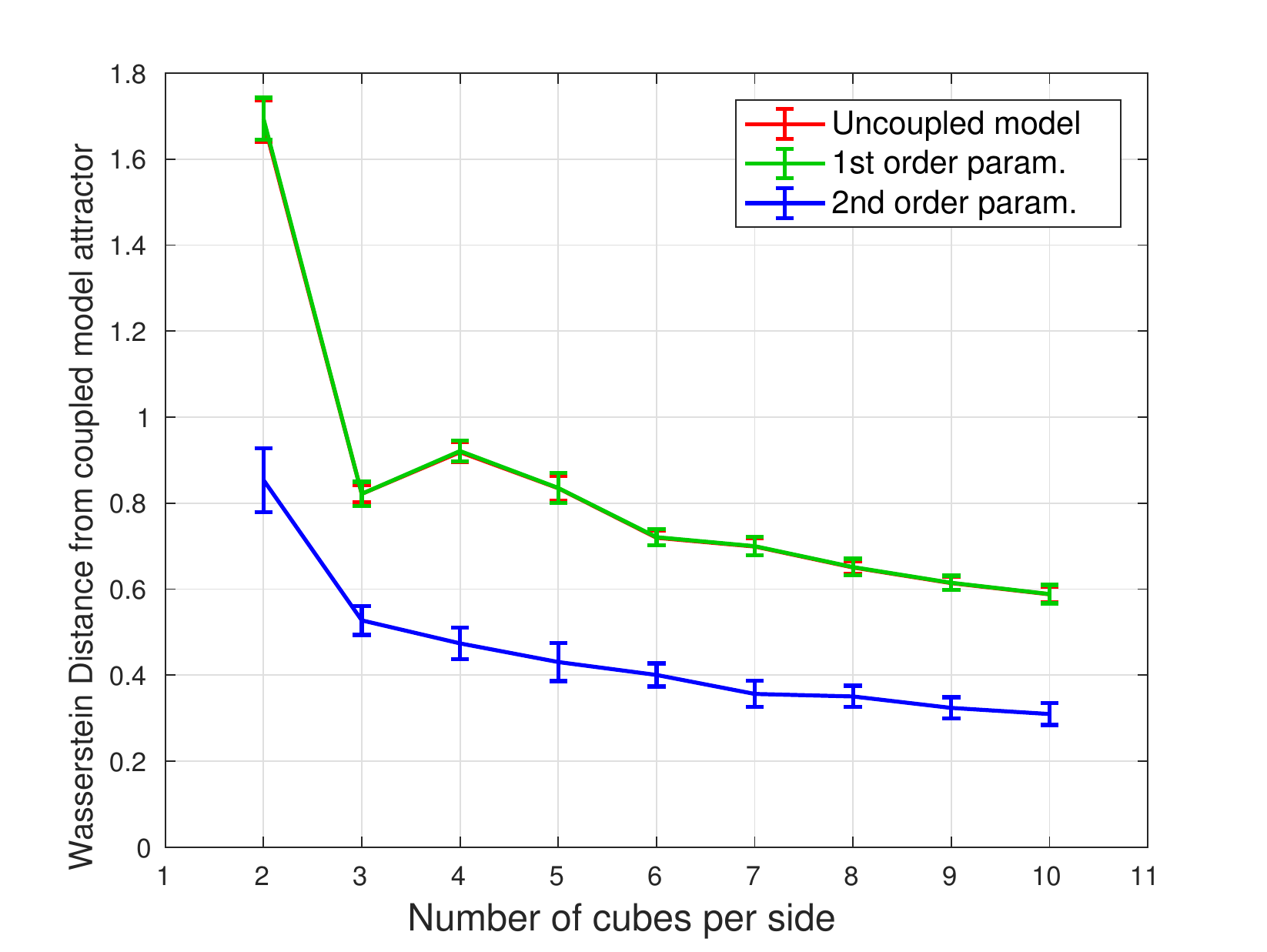}
  \caption{}
\end{subfigure}%
\begin{subfigure}{.5\textwidth}
  \includegraphics[width=\linewidth]{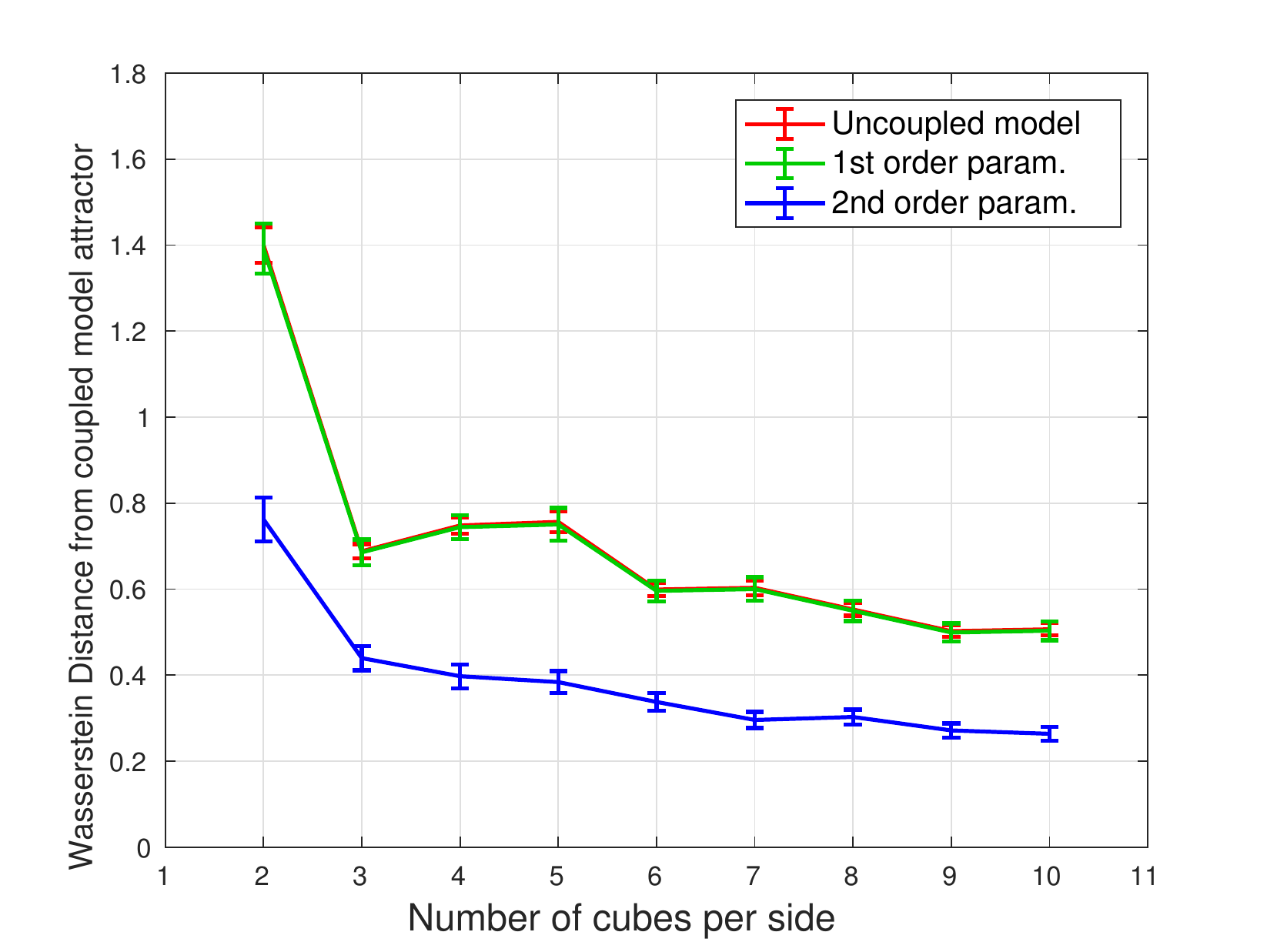}
  \caption{}
\end{subfigure}\\
\begin{subfigure}{.5\textwidth}
  \centering
  \includegraphics[width=\linewidth]{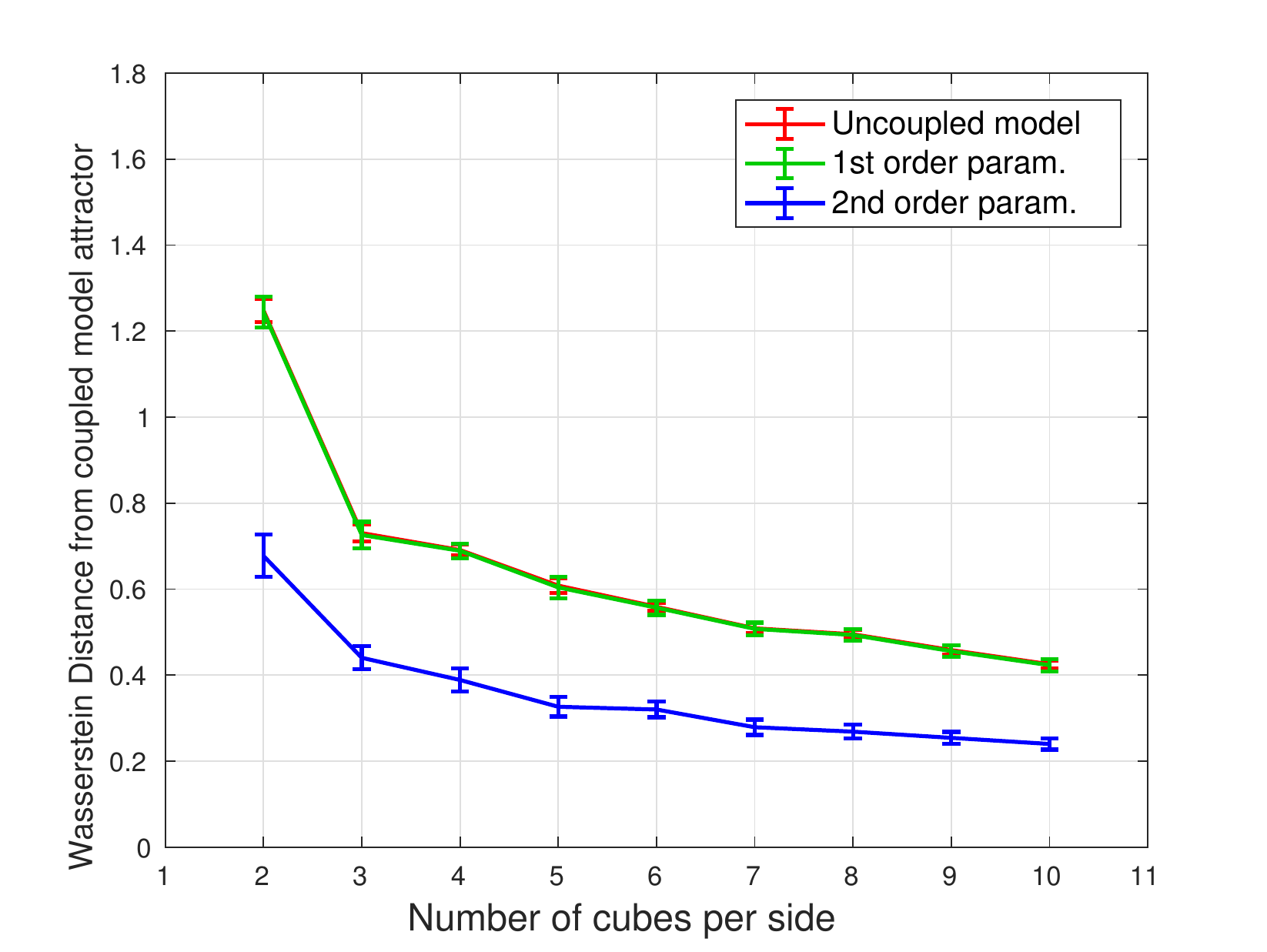}
  \caption{}
\end{subfigure}%
\begin{subfigure}{.5\textwidth}
  \includegraphics[width=\linewidth]{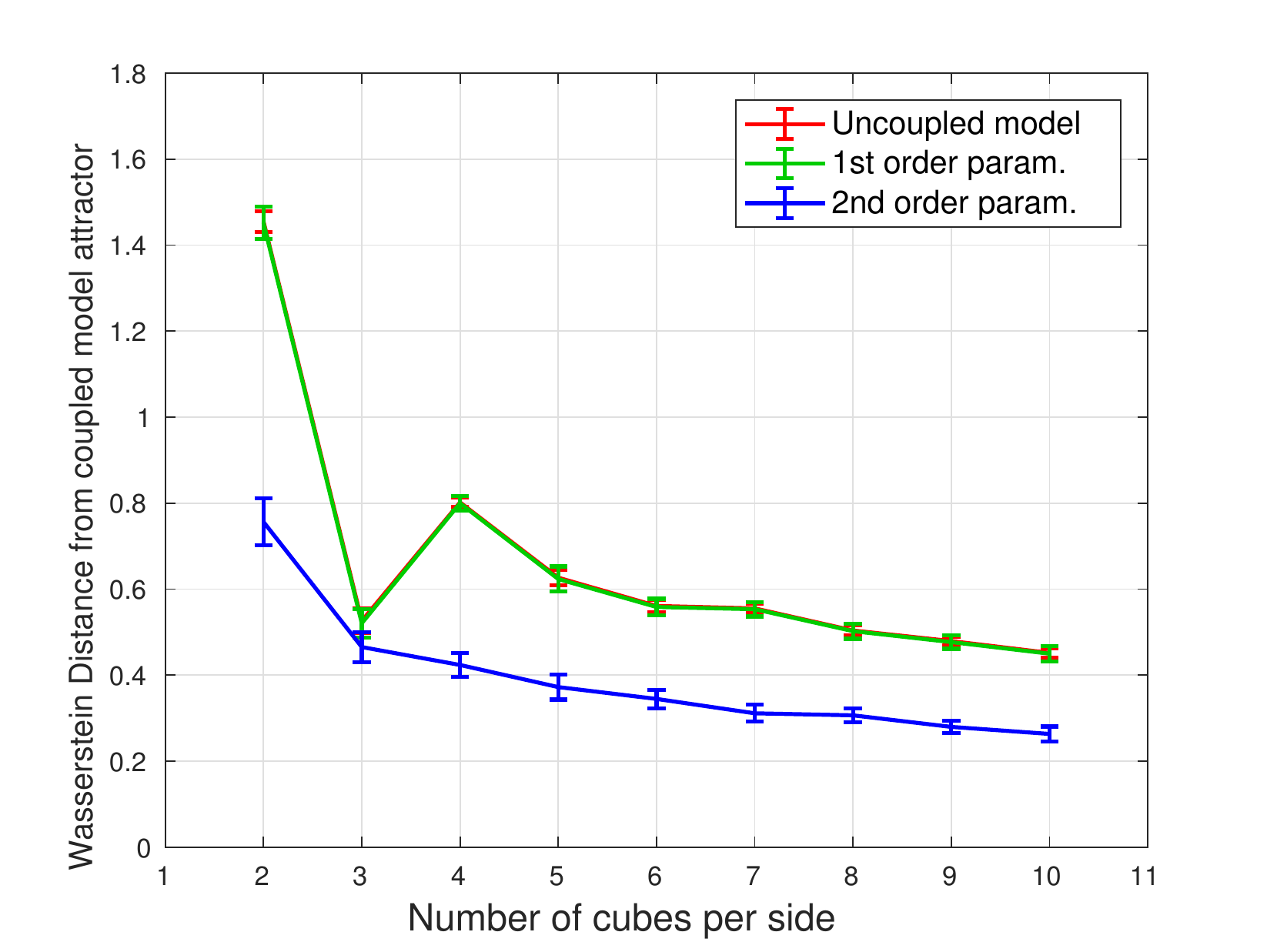}
  \caption{}
\end{subfigure}%
  \caption{Wasserstein distances from the coupled model with respect to number of cubes per side: a) $3D$ case; b) Projection on $XY$ plane; c) Projection on $XZ$ plane; d) Projection on $YZ$ plane.}
  \label{WD_n05}
\end{figure*}

The analysis performed considering the Wasserstein distance between the measures  is shown in Fig.\ref{WD_n05}. Without going into details, one finds that the same considerations we made for $\tau=5$ are still valid for $\tau=\frac{1}{6}$ regarding the performance of the parametrization schemes and the role of coarse graining. Additionally, we observe that, for each choice of coarse-graining, the distance between the measure of the parametrized models and the actual projected measure of the coupled model is larger for $\tau=\frac{1}{6}$, thus indicating the parametrization procedure performs worse in this case. This fits with the intuition one can have by checking out how well Panels b)-d) reproduce Panel a) in Fig. \ref{PoinSecZ0tau05} versus the case of Fig. \ref{PoinSecZ0}.


\section{Conclusions}
Developing parametrizations able to surrogate efficiently and accurately the dynamics of unresolved degrees of freedom is a central task in many areas of science, and especially in geosciences. There is no obvious protocol in testing parametrizations for complex systems, because one is bound to look only at specific observables of interests. This procedure is not error-free, because optimizing a parametrization against one or more observables might lead to unfortunate effects on other aspects of the system and worsen, in some other aspects, its performance.

In this paper we have addressed the problem of constructing a parametrization for a simple yet meaningful multiscale system, and then testing its performance in a possibly comprehensive way. We have considered a simple six-dimensional system constructed by coupling a Lorenz 84 system and a Lorenz 63 system, with the latter acting as forcing to the former, and the former being the subsystem of interest. We have included a parameter controlling the time scale separation of the two system and a parameter controlling the intensity of the coupling. We have built a first order and a second order parametrization able to surrogate the effects of the coupling using the scale-adaptive WL method. The second order scheme includes a stochastic term, which has proved to be essential for radically improving the quality of the parametrization with respect to the purely determinic case (first order parametrization), as already visually shown by looking at suitable Poincar\'e sections. 

We show here that, in agreement of what shown in previous papers, the WL-approach provides an accurate and flexible framework for constructing parametrizations. Nonetheless, the main novelty of this paper lies in our use of the Wasserstein distance as a comprehensive tool for measuring how different the invariant measures ("the climates") of the uncoupled Lorenz 84 model, and of its two version with deterministic and stochastic parametrizations are from the projection of the measure of the coupled model on the variables of the Lorenz 84 model. We discover that the Wasserstein distance provides a robust tool for assessing the quality of the parametrization, and, quite encouragingly, meaningful results can be obtained when considering very coarse grained representation of the phase space. A well-known issues of using a methodology like the Wasserstein distance is the so-called curse of dimensionality: the procedure itself becomes unfeasible when the system has a number of degree of freedom above few units. We have addressed (partially) this issue by looking at the Wasserstein distance of the projected measures on the three two-dimensional spaces spanned by two of the three variables of the Lorenz 84 model. We find that the properties of the Wasserstein distance in the reduced spaces follow closely those found in the full space. This might suggest as practice for testing parametrization the use of Wasserstein distance in suitably constructed low-dimensional projections of the full phase space. This has, in principle, much greater robustness than looking, e.g., at moments of the distributions and co-variances.

\section*{Acknowledgement}
The authors wish to thank G. Peyr\'e for making the Matlab software related to Wasserstein Distance publicly available. GV was supported by the Hans Ertel Center for Weather Research (HErZ), a collaborative project involving universities across Germany, the Deutscher Wetterdienst and funded by the BMVI (Federal Ministry of Transport and Digital Infrastructure, Germany). VL acknowledges the financial support provided by the DFG SFB/Transregio Project TRR181 and by the Horizon2020 projects BlueAction and CRESCENDO. VL wishes to thank M. Ghil for having suggested the relevance of the Wasserstein distance, and \cite{Robin2017} for having written a stimulating paper at this regard. VL recalls several fond memories of very informal yet enlightening discussions with A. Trevisan on nonlinear dynamics and data assimilation.


\end{document}